\documentclass[submission,copyright,creativecommons]{eptcs}
\usepackage[T1]{fontenc}
\usepackage{isabelle,isabellesym}

\usepackage{amsmath,amsthm}
\usepackage{mathpartir}
\usepackage{graphicx}
\usepackage{caption}
\usepackage{subcaption}

\newtheorem{definition}{Definition}
\newtheorem{lemma}{Lemma}
\newtheorem{theorem}{Theorem}
\newtheorem{corollary}{Corollary}

\usepackage[mode=buildnew]{standalone}

\usepackage{tikz}
\usetikzlibrary{arrows}

\isabellestyle{it}

\begin{document}

\title{Towards Mechanised Proofs in\\ Double-Pushout Graph Transformation}
\author{Robert S\"oldner  \qquad Detlef Plump
\institute{Department of Computer Science, University of York, York, UK
\email{\{rs2040,detlef.plump\}@york.ac.uk}}}
\def\authorrunning{R. S\"oldner and D. Plump}
\def\titlerunning{Towards Mechanised Proofs in Double-Pushout Graph Transformation}

\providecommand{\event}{GCM 2022}

\maketitle

\begin{abstract}
We formalise the basics of the double-pushout approach to graph transformation in the proof 
assistant Isabelle/HOL and provide associated machine-checked proofs.
Specifically, we formalise graphs, graph morphisms and rules, and a definition of direct derivations 
based on deletion and gluing. We then formalise graph pushouts and prove with Isabelle's help that 
both deletions and gluings are pushouts. We also prove that pushouts are unique up to isomorphism.
The formalisation comprises around 2000 lines of source text. Our motivation is to pave the way for
rigorous, machine-checked proofs in the theory of the double-pushout approach, and to lay the 
foundations for verifying graph transformation systems and rule-based graph programs by 
interactive theorem proving.

\end{abstract}


\parindent 0pt\parskip 0.5ex

\begin{isabellebody}%
\setisabellecontext{Content}%
\isadelimtheory
\endisadelimtheory
\isatagtheory
\endisatagtheory
{\isafoldtheory}%
\isadelimtheory
\endisadelimtheory
\isadelimdocument
\endisadelimdocument
\isatagdocument
\isamarkupsection{Introduction \label{sec:introduction}%
}
\isamarkuptrue%
\endisatagdocument
{\isafolddocument}%
\isadelimdocument
\endisadelimdocument
\begin{isamarkuptext}%
Software faults may lead to unexpected system's behaviour with a significant loss of goods or even 
personal harm. Documented examples of system failures range from medical devices \cite{Leveson93a}
over space launch vehicles \cite{Dowson97a} to hardware design \cite{Harrison03a}.
To prevent software faults,  formal methods such as static analysis or program verification continue
to attract a considerable amount of research.

Computing by rule-based graph transformation provides an intuitive and visual approach to 
specification and programming. Here, the main formal concepts for ensuring correctness are 
model checking \cite{Rensink-Schmidt-Varro04a,Varro04a,Baresi-Rafe-Rahmani-Spoletini08a,Rensink08a,Ghamarian-deMol-Rensink-Zambon-Zimakova12a} and proof-based verification
\cite{Habel-Pennemann09a,Inaba-Hidaka-Hu-Kato-Nakano11a,Poskitt-Plump14a,Stuckrath16a,Cavalheiro-Foss-Ribeiro17a,Strecker18a,Brenas18a,Wulandari-Plump21a}.
One of the oldest and most established approaches to graph transformation is the double-pushout (DPO)
approach, where rule applications are defined by a pair of pushouts in the category of graphs 
\cite{Ehrig-Ehrig-Prange-Taentzer06a}. Formal proofs in the DPO approach come in two flavours, 
they either establish results in the DPO theory (such as the commutativity of independent rule applications)
or they show the correctness of concrete graph transformation systems and graph programs.

While mainstream formal methods increasingly employ proof assistants such as Coq \cite{Bertot-Casteran04a} 
or Isabelle \cite{Nipkow-Klein14a} to obtain rigorous, machine-checked proofs, to the best of 
our knowledge such tools have not yet been used in the area of DPO graph transformation.
In this paper, we report on first steps towards the formalisation of the DPO theory in the 
Isabelle proof assistant. Specifically, we focus on linear rules with injective matching and show 
how to formalise (labelled, directed) graphs, morphisms, and rules. 
(Note that injective matching is more expressive than unrestricted matching because each rule can 
be replaced by the set of its quotient rules, and selected quotients can be omitted \cite{Habel-Mueller-Plump98a}).
We give an operational definition of direct derivations based on deletion and gluing. 
We then formalise graph pushouts and prove with Isabelle’s help that both deletion and 
gluing are pushouts. We also prove that pushouts are unique up to isomorphism.

We stress that we do not intend to formalise an abstract theoretical framework such as 
adhesive categories \cite{Ehrig-Ehrig-Prange-Taentzer06a}, nor do we aim at covering all kinds
of graphs that one can find in the DPO literature such as infinite graphs, hypergraphs, 
typed graphs, etc. Rather, we are interested in concrete constructions on graphs such as deletion 
and gluing, and how they relate to the double-pushout formulation. Our long-term goal is to provide
interactive and automatic proof support for formal reasoning on programs in a graph transformation
language such as GP\,2 \cite{Campbell-Courtehoute-Plump21a}.
The underlying formalisation in Isabelle will inevitably have to deal with the concrete graphs,
labels, rules, etc., which are the ingredients of such programs.

To summarize, this paper makes the following contributions:

\begin{itemize}%
\item We formalise in Isabelle the basics of the DPO approach with injective matching.

\item We prove that the operational construction of direct derivations by deletion and gluing
gives rise to a double-pushout diagram.    

\item We prove that graph pushouts are unique up to isomorphism.%
\end{itemize}

We believe that this is the first formalisation of DPO-based graph transformation in a theorem prover. 
The formalisation and proofs were developed using the Isabelle 2021 proof assistant. 
The entire formalisation comprises around 2000 lines of source text and can be accessed from 
GitHub\footnote{https://github.com/UoYCS-plasma/DPO-Formalisation}.

This paper is a revised version of \cite{Soeldner-Plump22a}. Here, we generalise our
formalisation to support gluing and deletion with injective morphisms.
Additionally, we follow Noschinski's~\cite{Noschinski2015} approach by using dedicated \isa{record} types
(for graphs and morphisms) and Isabelle's \isa{locale} mechanism.

The rest of the paper is structured as follows: Section \ref{sec:background} briefly reviews the 
theoretical background required in this research. Section \ref{sec:formalisation} will provide 
selected examples of our formalisation using the proof assistant Isabelle. 
Finally, in Section \ref{sec:conclusion}, the paper is summarised and future work is stated.%
\end{isamarkuptext}\isamarkuptrue%
\isadelimdocument
\endisadelimdocument
\isatagdocument
\isamarkupsection{Graphs, Rules and Derivations \label{sec:background}%
}
\isamarkuptrue%
\endisatagdocument
{\isafolddocument}%
\isadelimdocument
\endisadelimdocument
\begin{isamarkuptext}%
This section reviews basic terminology and results regarding graphs, rules, and derivations
  in the double-pushout approach with injective matching; see for example 
  \cite{Ehrig-Ehrig-Prange-Taentzer06a,Habel-Mueller-Plump98a}.
   In Section \ref{sec:formalisation}, we formalise these definitions and results in Isabelle. 
  
  \begin{definition}[Label alphabet]\label{def:alphabet}
  \normalfont
  A \emph{label alphabet} \(\mathcal{L} = (\mathcal{L}_V, \mathcal{L}_E)\) consists of a
  set \(\mathcal{L}_V\) of node labels and a set \(\mathcal{L}_E\) of edge labels.\qed
  \end{definition}
  
  We define directed and labelled graphs and allow parallel edges and loops.
  We do not consider variables as labels.
  \begin{definition}[Graph]\label{def:graph}
  \normalfont
  A \emph{graph} \(G = (V, E, s, t, l, m)\) over the alphabet \(\mathcal{L}\) is a system where 
  \(V\) is the finite set of nodes, \(E\) is the finite set of edges,
  \(s,t \colon E \to V\) functions assigning the source and target to each edge, 
  \(l \colon V \to \mathcal{L}_V\) and \(m \colon E \to \mathcal{L}_E\) are functions assigning
  a label to each node and edge.\qed
  \end{definition}

  Next we review graph morphisms which are structure-preserving mappings between graphs. 
  We describe our Isabelle formalisation in Subsection \ref{sec:graphs-morphisms}.

  \begin{definition}[Graph morphism]\label{def:morphism}
  \normalfont
  A \emph{graph morphism}\/ \(f \colon G \to H\) is a pair of mappings
  \(f = (f_V \colon V_G \to V_H,\, f_E \colon E_G \to E_H)\), such that
  for all \(e \in E_G\) and \(v \in V_G\):
  \begin{enumerate}
      \item \(f_V(s_G(e)) = s_H(f_E(e))\) (sources are preserved)
      \item \(f_V(t_G(e)) = t_H(f_E(e))\) (targets are preserved)
      \item \(l_G(v) = l_H(f_V(v))\) (node labels are preserved)
      \item \(m_G(e) = m_H(f_E(e))\) (edge labels are preserved)\qed
  \end{enumerate}
  \end{definition}
  
  We also define some special forms of morphisms.
  \begin{definition}[Special morphisms and isomorphic graphs]\label{def:special-morph}
  \normalfont
  A morphism \(f\) is \emph{injective}\/ (\emph{surjective}, \emph{bijective}) if \(f_V\) and
  \(f_E\) are injective (surjective, bijective). Morphism \(f\) is an \emph{inclusion} if 
  for all \(v \in V_G\)  and \(e \in V_E\), \(f_V(v) = v\) and \(f_E(e) = e\).
  A bijective morphism is an \emph{isomorphism}. In this case, \(G\) and \(H\) are
  \emph{isomorphic}, which is denoted by \(G \cong H\).\qed
  \end{definition}
  
  The composition of two morphisms yields a well-defined morphism, which 
  we prove in Subsection~\ref{sec:graphs-morphisms}.
  \begin{definition}[Morphism composition]\label{def:morphcomp}
  \normalfont
  Let \(f \colon F \rightarrow G\) and \(g \colon G \rightarrow H\) be graph morphisms. 
  The \emph{morphism composition}\/ \(g \circ f \colon F \rightarrow H\) is 
  defined by \(g \circ f = (g_V \circ f_V, g_E \circ f_E)\).\qed
  \end{definition}

  In DPO-based graph transformation, rules are the atomic units of computation. 
  We describe the formalisation of rules in Subsection~\ref{sec:rules-derivations}.

  \begin{definition}[Rule]\label{def:rule}
  \normalfont
  A \emph{rule}\/ \((L\leftarrow K \rightarrow R)\) consists of graphs \(L,K\) and \(R\) 
  over \(\mathcal{L}\) together with inclusions \(K \to L\) and \(K \to R\).\qed
  \end{definition}
  
  The addition of graph components along a common subgraph is called \emph{gluing}.
  We present our Isabelle formalisation in Subsection \ref{sec:gluing-pushouts}.
  The gluing construction below uses the disjoint union of sets $A$ and $B$ defined 
  by $A + B = (A \times \{1\}) \cup (B \times \{2\})$. It comes with injective functions
  $i_A \colon A \to A+B$ and $i_B \colon B \to A+B$ such that $i_A(A) \cup i_B(B) = A+B$ and 
  $i_A \cap i_B = \emptyset$.

  To keep the rest of this section readable, we tacitly assume that the injections $i_A$ and 
  $i_B$ are inclusions. Only in section \ref{sec:formalisation} we will be dealing 
  explicitly with the injections. We prove the correspondence between the gluing construction
  and pushouts in Subsection \ref{sec:gluing-pushouts}.
  
  \begin{lemma}[Gluing \cite{Ehrig79a}]\label{lemma:gluing}
  Let \(b \colon K \to R\) and \(d \colon K \to D\) be injective graph morphisms.
  Then the following defines a graph \(H\) (see Fig.~\ref{fig:gluing}), the gluing 
  of \(D\) and \(R\) according to \(d\):
  \begin{enumerate}
      \item $V_H = V_D + (V_R - b_V(V_K))$
      \item $E_H = E_D + (E_R - b_E(E_K))$
      \item \(s_H(e) = \begin{cases}
          s_D(e) & \textrm{if } e \in E_D  \\
          d_V(b_V^{-1}(s_R(e))) & \textrm{if } e \in E_R - b_E(E_K) \textrm{ and } s_R(e) \in b_V(V_K)  \\
          s_R(e) & \textrm{otherwise}
          \end{cases}
          \)
      \item $t_H$ analogous to $s_H$
      \item \(l_H = \begin{cases}
          l_D(v) & \textrm{if } v \in V_D \\
          l_R(v) & \textrm{otherwise}
          \end{cases}\)
      \item $m_H$ analogous to $l_H$
  \end{enumerate}
  Moreover, the morphism $D \to H$ is an inclusion and the injective morphism $h$ is defined for 
  all items $x$ in $R$ by $h(x) = \textbf{if } x \in R - b(K) \textbf{ then } x \textbf{ else } d(x)$.
  \end{lemma}
  
  The dangling condition ensures that deletion results in a well-defined graph.
  \begin{definition}[Dangling condition]\label{def:dang}
  \normalfont
  Let \(b' \colon K \to L\) be an injective graph morphism. 
  An injective graph morphism \(g \colon L \to G\) satisfies the 
  \emph{dangling condition}\/ if no edge in $E_G - g_E(E_L)$ is 
  incident to a node in $g_V(V_L - b'_V(V_K))$.\qed
  \end{definition}
  
  The following \emph{deletion} of graph components is  formalised in Subsection \ref{sec:deletion-pushouts}.

  \begin{lemma}[Deletion \cite{Ehrig79a}]\label{lemma:deletion}
  Let \(b' \colon K \to L\) and  $g \colon L \to G$ injective graph morphisms and
  let \(g\) satisfy the dangling condition (see Fig.~\ref{fig:deletion}). Then
  the following defines a graph \(D\), the deletion of \(L\) and \(G\) according to \(d\).
  \begin{enumerate}
      \item \(V_D = V_G - g_V(V_L - b'_V(V_K))\) and \(E_D = E_G - g_E(E_L - b'_E(E_K))\) induce the inclusion \(D \to G\), and
      \item there is an injective graph morphism $d \colon K \rightarrow D$, 
        defined by $d(x) = g(b'(x))$ for all items $x$ in $K$.
  \end{enumerate}
  \end{lemma}

  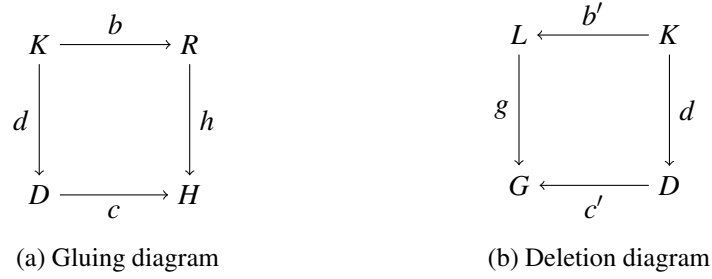
\begin{figure}
      \centering
      \begin{subfigure}[b]{0.4\textwidth}
          \centering
          \begin{tikzpicture}[node distance=2cm, auto]
  \node (K) {$K$};
  \node[right of=K] (R) {$R$};
  \node[below of=K] (D) {$D$};
  \node[below of=R] (H) {$H$};

  \draw[->] (K) to node{$b$} (R);
  \draw[->] (K) to node[swap]{$d$} (D);
  \draw[->] (R) to node{$h$} (H);
  \draw[->] (D) to node[swap]{$c$} (H);
\end{tikzpicture}
          \caption{Gluing diagram}
          \label{fig:gluing}
      \end{subfigure}%
      \begin{subfigure}[b]{0.4\textwidth}
          \centering
          \begin{tikzpicture}[node distance=2cm, auto]
  \node (K) {$K$};
  \node[left of=K] (L) {$L$};
  \node[below of=K] (D) {$D$};
  \node[below of=L] (G) {$G$};
  
  \draw[->] (K) to node[swap]{$b'$} (L);
  \draw[->] (K) to node{$d$} (D);
  \draw[->] (L) to node[swap]{$g$} (G);
  \draw[->] (D) to node{$c'$}(G);
\end{tikzpicture}
          \caption{Deletion diagram}
          
          \label{fig:deletion}
      \end{subfigure}
      \caption{Gluing and deletion diagram}
      \label{fig:deletion-guing}
  \end{figure}
  The following definition introduces the concept of \emph{pushouts} in the category of graphs.
  \begin{definition}[Pushout]\label{def:pushout}
  \normalfont
  Given graph morphisms $b \colon A \to B$ and $c \colon A \to C$, a graph $D$ together with 
  graph morphisms $f \colon B \to D$ and $g \colon C \to D$ is a \emph{pushout} of 
  $A \to B$ and $A \to C$ if the following holds (see Fig.~\ref{fig:pushout}):
  \begin{enumerate}
      \item Commutativity: $f \circ b = g \circ c$
      \item Universal property: For all graph morphisms $p \colon B \to H$ and $t \colon C \to H$ 
        such that $p \circ b = t \circ c$, there is a unique morphism $u \colon D \to H$ 
        such that $u \circ f = p$ and $u \circ g = t$.\qed
  \end{enumerate}
  \end{definition}
  
  The formalisation of pushouts and the proof that pushouts are unique up to isomorphism is 
  presented in Subsection \ref{sec:pushouts}. 
  \begin{theorem}[Uniqueness of pushouts \cite{AHS90}]\label{theorem:po-uniqueness}
  Let $A \to B$ and $A \to C$ together with $D$ induce a pushout as depicted in 
  Fig.~\ref{fig:pushout}. A graph $H$ together with morphisms $B \to H$ and $C \to H$ is a pushout
  of $b$ and $c$ if and only if there is an isomorphism $u \colon D \to H$ such that $u \circ f = p$
  and $u \circ g = t$.
  \end{theorem}
  \begin{figure}
      \centering
      \begin{minipage}[t]{0.4\textwidth}%
          \begin{tikzpicture}[node distance=2cm, auto]
  \node (A) {$A$};
  \node[right of=A] (B) {$B$};
  \node[below of=A] (C) {$C$};
  \node[below of=B] (D) {$D$};
  \node[node distance=1cm, right of=D, below of=D] (H) {$H$};

  \draw[->] (A) to node{$b$} (B);
  \draw[->] (A) to node[swap]{$c$} (C);
  \draw[->] (B) to node{$f$} (D);
  \draw[->] (C) to node[swap]{$g$} (D);

  \draw[->, bend left ] (B) to node{$p$} (H);
  \draw[->, bend right] (C) to node[swap]{$t$} (H);

  \draw[->, dashed] (D) to node{$u$} (H);

\end{tikzpicture}
          \caption{Pushout diagram}
          \label{fig:pushout}
      \end{minipage}\hfill
      \begin{minipage}[t]{0.4\textwidth}%
          \begin{tikzpicture}[node distance=2cm, auto]

    \node (L) {L};
    \node[right of=L] (K) {$K$};
    \node[right of=K] (R) {$R$};
    
    \node[below of=L] (G) {$G$};
    \node[below of=K] (D) {$D$};
    \node[below of=R] (H) {$H$};
    
    \node[node distance=0.5cm, right of=H] (CC) {$\cong$};
    \node[node distance=0.5cm, right of=CC] {$M$};
    
    \node[node distance=1cm, right of=L, below of=L] {(1)};
    \node[node distance=1cm, right of=K, below of=K] {(2)};
    
    \draw[->] (K) to (L) {};
    \draw[->] (K) to (R) {};
    \draw[->] (L) to (G) {};
    \draw[->] (D) to (G) {};
    \draw[->] (D) to (H) {};
    \draw[->] (R) to (H) {};
    \draw[->] (K) to (D) {};

\end{tikzpicture}
          \caption{Direct derivation}
          \label{fig:direct-derivation}
      \end{minipage}%
  \end{figure}
  
  \begin{theorem}[Gluings are pushouts \cite{Ehrig79a}]\label{theorem:gluing-po}
  Let \(b \colon K \rightarrow R\) and \(d \colon K \rightarrow D\) be injective graph morphisms,
  and \(H\) be the gluing of \(D\) and \(R\) according to \(d\), as defined in Lemma~\ref{lemma:gluing}. 
  Then, the square in Fig.~\ref{fig:gluing} is a pushout diagram where \(D \to H\) is an inclusion
  and \(h\) is defined by \(h(x) = \textbf{if } x \in R - b(K) \textbf{ then } x \textbf{ else } c(d(x))\).
  We call \(H\) the pushout object.
  \end{theorem}
  
  The deletion construction of Lemma~\ref{lemma:deletion} and the following theorem are formalised
  and proved in Subsection~\ref{sec:deletion-pushouts}.
  
  \begin{theorem}[Deletions are pushouts \cite{Ehrig79a}]\label{theorem:deletion-po}
  Let \(K \rightarrow L\) and $g \colon L \to G$ be injective graph morphisms 
  and let \(g\) satisfy the dangling condition and the subgraph \(D\) of \(G\) as defined in 
  Lemma~\ref{lemma:deletion}. Then, the square in Fig.~\ref{fig:deletion} is a pushout diagram 
  where \(g\) is an inclusion and \(d(x) = g(b'(x))\) for all items \(x\) in \(K\).
  We call \(D\) the pushout complement.
  \end{theorem}
  
  The following definition of rule application is formalised in Subsection \ref{sec:rules-derivations}.
  
  \begin{definition}[Direct derivation]\label{def:direct-derivation}
  \normalfont
  Let $r = (L\leftarrow K \rightarrow R)$ be a rule and $g \colon L \to G$ be an injective 
  graph morphism satisfying the dangling condition. 
  Then $G$ directly derives (see Fig.~\ref{fig:direct-derivation}) $M$ by $r$ and $g$,
  denoted by $G \Rightarrow_{r,g} M$, if $H \cong M$, where $H$ is constructed from $G$ by:
  \begin{enumerate}
      \item (Deletion) $D$ is the subgraph $G - g(L - b'(K))$.
      \item $d \colon K \rightarrow D$ is the restriction of $g$ to $K$ and $D$.
      \item (Gluing) $H$ is the gluing $H = D + (R - b(K))$. \qed
  \end{enumerate}
  \end{definition}
  
  The following corollary follows directly by Theorem 2 and Theorem 3.

  \begin{corollary}[Direct derivation are double-pushouts \label{corollary:dd-po}]
   Given a direct derivation $G \Rightarrow_{r,g} M$, squares (1) and (2) in Figure~\ref{fig:direct-derivation} are pushouts. 
  \end{corollary}

  The next section provides a general introduction to the Isabelle proof assistant and 
  highlights selected parts of our formalisation.%
\end{isamarkuptext}\isamarkuptrue%
\isadelimdocument
\endisadelimdocument
\isatagdocument
\isamarkupsection{DPO Formalisation in Isabelle/HOL \label{sec:formalisation}%
}
\isamarkuptrue%
\endisatagdocument
{\isafolddocument}%
\isadelimdocument
\endisadelimdocument
\begin{isamarkuptext}%
Isabelle is a generic, interactive theorem prover based on the so-called \emph{LCF} approach.
  It is based on a small (meta-logical) proof kernel, which is responsible for checking all proofs.
  This concept provides high confidence in the prover's soundness. Isabelle/HOL refers to the 
  higher-order logic instantiation which is considered to be the most established calculus within
  the Isabelle distribution \cite{Paulson2019}.
  
  In Isabelle, type variables are denoted by a leading apostrophe. A term \isa{f} of type \isa{{\isacharprime}{\kern0pt}a}
  is denoted by \isa{f{\isacharcolon}{\kern0pt}{\isacharcolon}{\kern0pt}{\isacharprime}{\kern0pt}a}. The function type from \isa{{\isacharprime}{\kern0pt}a} to \isa{{\isacharprime}{\kern0pt}b} is written
  \isa{f{\isacharcolon}{\kern0pt}{\isacharcolon}{\kern0pt}{\isacharprime}{\kern0pt}a\ {\isasymRightarrow}\ {\isacharprime}{\kern0pt}b}. 
  The inference rule notation \isa{{\isasymlbrakk}\ A\isactrlsub {\isadigit{1}}{\isacharsemicolon}{\kern0pt}\ A\isactrlsub {\isadigit{2}}\ {\isasymrbrakk}\ {\isasymLongrightarrow}\ C} with premises \isa{A\isactrlsub {\isadigit{1}}} and \isa{A\isactrlsub {\isadigit{2}}} and conclusion \isa{C} is 
  a shorthand (with the \isa{{\isacharsemicolon}{\kern0pt}} (semicolon) as a logical \emph{and}) for the implication \isa{A\isactrlsub {\isadigit{1}}\ {\isasymLongrightarrow}\ A\isactrlsub {\isadigit{2}}\ {\isasymLongrightarrow}\ C}.
  Its natural representation is given by:
  \begin{center}
    \isa{\mbox{}\inferrule{\mbox{A\isactrlsub {\isadigit{1}}}\\\ \mbox{A\isactrlsub {\isadigit{2}}}}{\mbox{C}}}
  \end{center}
  Isabelle meta-logical universal quantifier \isa{{\isasymAnd}} corresponds to HOL's \isa{{\isasymforall}} and the meta implication
  \isa{{\isasymLongrightarrow}} to \isa{{\isasymlongrightarrow}}. The meta logic is used to expressed inference rules and cannot appear in 
  HOL formulae.

  Our formalisation is based on Isabelle's \isacommand{locale} mechanism, a technique for writing
  parametric specifications. Furthermore, we use \emph{intelligible semi-automated reasoning} (Isar) 
  which is Isabelle's language of writing structured proofs \cite{Wenzel1999}.
  In contrast to \emph{apply-scripts}, which execute deduction rules in a linear manner, Isar follows 
  a structured approach resulting in increased readability and maintainability \cite{Nipkow-Klein14a}.
  
  A general introduction to Isabelle/HOL can be found in \cite{Nipkow-Klein14a}.
  The main components of our formalisation and their interdependencies are depicted in 
  Fig.\ref{fig:structure}. The simple arrow ($\rightarrow$) can be read as "depends on", i.e., the 
  definition of morphisms depends on the definition of graphs allowing the inheritance of properties.
  The blue arrow  \textcolor{blue}{$\Longrightarrow$} highlights main theorems proven in this 
  study, viz.\ that the gluing and deletion constructions correspond to pushouts and that 
  pushout objects are unique up to isomorphism.

  \begin{figure}
      \centering
      \usetikzlibrary{arrows}

\begin{tikzpicture}   
    \node (G) at (0,1) {Graph};
    \node (M) at (0,2) {Morphism};
    \node (Gl) at (2.5, 4) {Gluing};
    \node (D) at (-2.5, 4) {Deletion};
    \node (P) at (0, 4) {Pushout};
    \node (U) at (0, 5) {Uniqueness};
    \node (Di) at (0, 6) {Direct Derivation};
    
    \draw[->] (M) to (G) {};
    \draw [<-, bend right] (M.east) to  (Gl.south) {};
    \draw [<-, bend left] (M.west) to  (D.south) {};
    \draw [<-, bend left] (D.north) to  (Di.west) {};
    \draw [<-, bend right] (Gl.north) to  (Di.east) {};
    \draw [->] (P) to  (M) {};

   \draw[-implies,double equal sign distance, blue] (Gl) to (P){};
   \draw[-implies,double equal sign distance, blue] (D) to (P){};
   \draw[-implies,double equal sign distance, blue] (P) to (U){};  
\end{tikzpicture}
      \caption{Overview of component dependencies ($\to$) and major theorems (\textcolor{blue}{$\Rightarrow$})}
      \label{fig:structure}
  \end{figure}
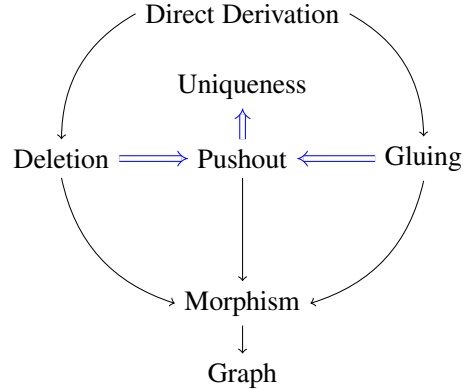

  The upcoming subsection introduces the basic building blocks of our formalisation:
  Graphs and graph morphisms.%
\end{isamarkuptext}\isamarkuptrue%
\isadelimdocument
\endisadelimdocument
\isatagdocument
\isamarkupsubsection{Graphs and Morphisms\label{sec:graphs-morphisms}%
}
\isamarkuptrue%
\endisatagdocument
{\isafolddocument}%
\isadelimdocument
\endisadelimdocument
\begin{isamarkuptext}%
Our definition of graphs (Def.~\ref{def:graph}) is different from Strecker's \cite{Strecker18a} 
  where a graph is a set of nodes together with a binary relation of nodes. 
  A consequence of Strecker's definition is the absence of parallel edges and edge labels.
  We follow Noschinski's graph library \cite{Noschinski2015} approach in the sense, that we use
  a \isacommand{record} data structure to represent a graph and enforce the well-formedness by using
  the \isacommand{locale} mechanism. We extend Noschinski's data structure to carry node 
  and edge labelling functions.

  The usage of type variables for node and edge identifiers (\isa{{\isacharprime}{\kern0pt}v} and \isa{{\isacharprime}{\kern0pt}e}), and labels
  (\isa{{\isacharprime}{\kern0pt}l} and \isa{{\isacharprime}{\kern0pt}m}) allows us to reason about an arbitrary representation. Common examples 
  include natural numbers (\isa{nat}) and strings (\isa{string}).

  We define graphs using the \isacommand{record} keyword as follows: 
\begin{isabelle}%
\isacommand{record}\ {\isacharparenleft}{\kern0pt}{\isacharprime}{\kern0pt}v{\isacharcomma}{\kern0pt}{\isacharprime}{\kern0pt}e{\isacharcomma}{\kern0pt}{\isacharprime}{\kern0pt}l{\isacharcomma}{\kern0pt}{\isacharprime}{\kern0pt}m{\isacharparenright}{\kern0pt}\ pre{\isacharunderscore}{\kern0pt}graph\ {\isacharequal}{\kern0pt}\isanewline
\ \ \ \ \ \ nodes\ \ {\isacharcolon}{\kern0pt}{\isacharcolon}{\kern0pt}\ {\isachardoublequoteopen}{\isacharprime}{\kern0pt}v\ set{\isachardoublequoteclose}\isanewline
\ \ \ \ \ \ edges\ \ {\isacharcolon}{\kern0pt}{\isacharcolon}{\kern0pt}\ {\isachardoublequoteopen}{\isacharprime}{\kern0pt}e\ set{\isachardoublequoteclose}\isanewline
\ \ \ \ \ \ source\ {\isacharcolon}{\kern0pt}{\isacharcolon}{\kern0pt}\ {\isachardoublequoteopen}{\isacharprime}{\kern0pt}e\ {\isasymRightarrow}\ {\isacharprime}{\kern0pt}v{\isachardoublequoteclose}\isanewline
\ \ \ \ \ \ target\ {\isacharcolon}{\kern0pt}{\isacharcolon}{\kern0pt}\ {\isachardoublequoteopen}{\isacharprime}{\kern0pt}e\ {\isasymRightarrow}\ {\isacharprime}{\kern0pt}v{\isachardoublequoteclose}\isanewline
\ \ \ \ \ \ node{\isacharunderscore}{\kern0pt}label\ {\isacharcolon}{\kern0pt}{\isacharcolon}{\kern0pt}\ {\isachardoublequoteopen}{\isacharprime}{\kern0pt}v\ {\isasymRightarrow}\ {\isacharprime}{\kern0pt}l{\isachardoublequoteclose}\isanewline
\ \ \ \ \ \ edge{\isacharunderscore}{\kern0pt}label\ {\isacharcolon}{\kern0pt}{\isacharcolon}{\kern0pt}\ {\isachardoublequoteopen}{\isacharprime}{\kern0pt}e\ {\isasymRightarrow}\ {\isacharprime}{\kern0pt}m{\isachardoublequoteclose}%
\end{isabelle}

  With the \isacommand{abbreviation} command, term abbreviations are introduced. 
  The (built-in) axiomatized term \isa{undefined{\isacharcolon}{\kern0pt}{\isacharcolon}{\kern0pt}{\isacharprime}{\kern0pt}a} is used to refer to a fixed but arbitrary
  term of type \isa{{\isacharprime}{\kern0pt}a}.

  Following this, we can define an abbreviation \isa{G}, representing the empty graph (\isa{pre{\isacharunderscore}{\kern0pt}graph}) 
  structure  as follows:%
\end{isamarkuptext}\isamarkuptrue%
\isacommand{abbreviation}\isamarkupfalse%
\ G\ \isakeyword{where}\isanewline
{\isacartoucheopen}G\ {\isasymequiv}\ {\isasymlparr}nodes\ {\isacharequal}{\kern0pt}\ {\isacharbraceleft}{\kern0pt}{\isacharbraceright}{\kern0pt}{\isacharcomma}{\kern0pt}\ edges\ {\isacharequal}{\kern0pt}\ {\isacharbraceleft}{\kern0pt}{\isacharbraceright}{\kern0pt}{\isacharcomma}{\kern0pt}\ source{\isacharequal}{\kern0pt}undefined{\isacharcomma}{\kern0pt}\ target{\isacharequal}{\kern0pt}undefined\isanewline
\ \ \ \ \ {\isacharcomma}{\kern0pt}node{\isacharunderscore}{\kern0pt}label{\isacharequal}{\kern0pt}undefined{\isacharcomma}{\kern0pt}\ edge{\isacharunderscore}{\kern0pt}label{\isacharequal}{\kern0pt}undefined{\isasymrparr}{\isacartoucheclose}%
\begin{isamarkuptext}%
We introduce a notation for working with the \isa{pre{\isacharunderscore}{\kern0pt}graph} structure, 
closely following Definition~\ref{def:graph}, using Isabelle's \isacommand{notation} keyword.
Throughout the formalization, we use \isa{V\isactrlbsub {\isacharunderscore}{\kern0pt}\isactrlesub } and \isa{E\isactrlbsub {\isacharunderscore}{\kern0pt}\isactrlesub } to refer to the set of nodes and edges,
\isa{s\isactrlbsub {\isacharunderscore}{\kern0pt}\isactrlesub } and \isa{t\isactrlbsub {\isacharunderscore}{\kern0pt}\isactrlesub } to refer to source and target functions, and \isa{l\isactrlbsub {\isacharunderscore}{\kern0pt}\isactrlesub } and \isa{m\isactrlbsub {\isacharunderscore}{\kern0pt}\isactrlesub } refer to the
node-label an edge-label functions. This allows us, for example, to write \isa{V\isactrlbsub G\isactrlesub } instead of 
\isa{nodes\ G} to refer to the set of nodes of a graph \(G\).

As the \isa{pre{\isacharunderscore}{\kern0pt}graph} record does not introduce any constraints, not only well-formed graphs can
be represented but also ill-formed graphs such as a graph with edges but no nodes.
The well-formedness is enforced via the \isa{graph} locale. Here, the \isa{fixes} keyword is used
to declare parameters while the \isa{assumes} keyword is used to state premises which hold within the
locale context. The \isa{graph} locale is defined as follows:

\begin{isabelle}%
\isacommand{locale}\ graph\ {\isacharequal}{\kern0pt}\isanewline
\ \ \ \ \ \ \isakeyword{fixes}\ G\ {\isacharcolon}{\kern0pt}{\isacharcolon}{\kern0pt}\ {\isachardoublequoteopen}{\isacharparenleft}{\kern0pt}{\isacharprime}{\kern0pt}v{\isacharcomma}{\kern0pt}{\isacharprime}{\kern0pt}e{\isacharcomma}{\kern0pt}{\isacharprime}{\kern0pt}l{\isacharcomma}{\kern0pt}{\isacharprime}{\kern0pt}m{\isacharparenright}{\kern0pt}\ pre{\isacharunderscore}{\kern0pt}graph{\isachardoublequoteclose}\isanewline
\ \ \ \ \ \ \isakeyword{assumes}\ \isanewline
\ \ \ \ \ \ \ \ finite{\isacharunderscore}{\kern0pt}nodes{\isacharcolon}{\kern0pt}\ {\isachardoublequoteopen}finite\ V\isactrlbsub G\isactrlesub {\isachardoublequoteclose}\ \isakeyword{and}\isanewline
\ \ \ \ \ \ \ \ finite{\isacharunderscore}{\kern0pt}edges{\isacharcolon}{\kern0pt}\ {\isachardoublequoteopen}finite\ E\isactrlbsub G\isactrlesub {\isachardoublequoteclose}\ \isakeyword{and}\isanewline
\ \ \ \ \ \ \ \ source{\isacharunderscore}{\kern0pt}integrity{\isacharcolon}{\kern0pt}\ {\isachardoublequoteopen}e\ {\isasymin}\ E\isactrlbsub G\isactrlesub \ {\isasymLongrightarrow}\ s\isactrlbsub G\isactrlesub \ e\ {\isasymin}\ V\isactrlbsub G\isactrlesub {\isachardoublequoteclose}\ \isakeyword{and}\isanewline
\ \ \ \ \ \ \ \ target{\isacharunderscore}{\kern0pt}integrity{\isacharcolon}{\kern0pt}\ {\isachardoublequoteopen}e\ {\isasymin}\ E\isactrlbsub G\isactrlesub \ {\isasymLongrightarrow}\ t\isactrlbsub G\isactrlesub \ e\ {\isasymin}\ V\isactrlbsub G\isactrlesub {\isachardoublequoteclose}%
\end{isabelle}

In this formalisation, the premises are:

\begin{itemize}%
\item The set of nodes (\isa{finite{\isacharunderscore}{\kern0pt}nodes}) and edges (\isa{finite{\isacharunderscore}{\kern0pt}edges}) are finite,

\item and the source (\isa{source{\isacharunderscore}{\kern0pt}integrity}) and target (\isa{target{\isacharunderscore}{\kern0pt}integrity}) functions map each edge to 
a node within the graph.%
\end{itemize}

We do not have to state explicit premises for both, node and edge, labelling functions as they 
are defined for the entire universe of the corresponding type (\isa{{\isacharprime}{\kern0pt}v} and \isa{{\isacharprime}{\kern0pt}e}).

We can prove, the \isa{pre{\isacharunderscore}{\kern0pt}graph} structure \isa{G} is indeed a \isa{graph} according to 
our \isacommand{locale} definition by using the \isacommand{interpretation} command as follows:%
\end{isamarkuptext}\isamarkuptrue%
\isacommand{interpretation}\isamarkupfalse%
\ graph\ G\isanewline
\isadelimproof
\ \ %
\endisadelimproof
\isatagproof
\isacommand{by}\isamarkupfalse%
\ unfold{\isacharunderscore}{\kern0pt}locales\ simp{\isacharunderscore}{\kern0pt}all%
\endisatagproof
{\isafoldproof}%
\isadelimproof
\endisadelimproof
\begin{isamarkuptext}%
The \isa{unfold{\isacharunderscore}{\kern0pt}locales} tactic applies all introduction rules generated by the 
  \isacommand{locale} command to the current proof goal. The introduction rule for the \isa{graph}
  locale is given by:

  \begin{center}
    \isa{\mbox{}\inferrule{\mbox{finite\ V\isactrlbsub G\isactrlesub }\\\ \mbox{finite\ E\isactrlbsub G\isactrlesub }\\\ \mbox{{\isasymAnd}e{\isachardot}{\kern0pt}\ \mbox{}\inferrule{\mbox{e\ {\isasymin}\ E\isactrlbsub G\isactrlesub }}{\mbox{s\isactrlbsub G\isactrlesub \ e\ {\isasymin}\ V\isactrlbsub G\isactrlesub }}}\\\ \mbox{{\isasymAnd}e{\isachardot}{\kern0pt}\ \mbox{}\inferrule{\mbox{e\ {\isasymin}\ E\isactrlbsub G\isactrlesub }}{\mbox{t\isactrlbsub G\isactrlesub \ e\ {\isasymin}\ V\isactrlbsub G\isactrlesub }}}}{\mbox{graph\ G}}}
  \end{center}

  To prove our graph structure \isa{{\isasymlparr}nodes\ {\isacharequal}{\kern0pt}\ {\isasymemptyset}{\isacharcomma}{\kern0pt}\ edges\ {\isacharequal}{\kern0pt}\ {\isasymemptyset}{\isacharcomma}{\kern0pt}\ source\ {\isacharequal}{\kern0pt}\ undefined{\isacharcomma}{\kern0pt}\ target\ {\isacharequal}{\kern0pt}\ undefined{\isacharcomma}{\kern0pt}\ node{\isacharunderscore}{\kern0pt}label\ {\isacharequal}{\kern0pt}\ undefined{\isacharcomma}{\kern0pt}\ edge{\isacharunderscore}{\kern0pt}label\ {\isacharequal}{\kern0pt}\ undefined{\isasymrparr}{\isacharcolon}{\kern0pt}{\isacharcolon}{\kern0pt}{\isacharparenleft}{\kern0pt}{\isacharprime}{\kern0pt}a{\isacharcomma}{\kern0pt}\ {\isacharprime}{\kern0pt}b{\isacharcomma}{\kern0pt}\ {\isacharprime}{\kern0pt}c{\isacharcomma}{\kern0pt}\ {\isacharprime}{\kern0pt}d{\isacharparenright}{\kern0pt}\ pre{\isacharunderscore}{\kern0pt}graph} fulfills the \isa{graph} premises, we have to
  prove, the set of nodes (edges) is finite and the source (target) integrity.
  Isabelle's simplifier is able to discharge these goals automatically.%
\end{isamarkuptext}\isamarkuptrue%
\begin{isamarkuptext}%
Our definition of graph morphisms (cf. Def.~\ref{def:morphism}) follows a similar pattern.
  We define the graph morphism data structure (\isa{{\isacharparenleft}{\kern0pt}{\isacharprime}{\kern0pt}v\isactrlsub {\isadigit{1}}{\isacharcomma}{\kern0pt}\ {\isacharprime}{\kern0pt}v\isactrlsub {\isadigit{2}}{\isacharcomma}{\kern0pt}\ {\isacharprime}{\kern0pt}e\isactrlsub {\isadigit{1}}{\isacharcomma}{\kern0pt}\ {\isacharprime}{\kern0pt}e\isactrlsub {\isadigit{2}}{\isacharparenright}{\kern0pt}\ pre{\isacharunderscore}{\kern0pt}graph}) with a 
  dedicated function for the nodes and edges:

\begin{isabelle}%
\isacommand{record}\ {\isacharparenleft}{\kern0pt}{\isacharprime}{\kern0pt}v\isactrlsub {\isadigit{1}}{\isacharcomma}{\kern0pt}{\isacharprime}{\kern0pt}v\isactrlsub {\isadigit{2}}{\isacharcomma}{\kern0pt}{\isacharprime}{\kern0pt}e\isactrlsub {\isadigit{1}}{\isacharcomma}{\kern0pt}{\isacharprime}{\kern0pt}e\isactrlsub {\isadigit{2}}{\isacharparenright}{\kern0pt}\ pre{\isacharunderscore}{\kern0pt}morph\ {\isacharequal}{\kern0pt}\isanewline
\ \ \ \ node{\isacharunderscore}{\kern0pt}map\ {\isacharcolon}{\kern0pt}{\isacharcolon}{\kern0pt}\ {\isachardoublequoteopen}{\isacharprime}{\kern0pt}v\isactrlsub {\isadigit{1}}\ {\isasymRightarrow}\ {\isacharprime}{\kern0pt}v\isactrlsub {\isadigit{2}}{\isachardoublequoteclose}\isanewline
\ \ \ \ edge{\isacharunderscore}{\kern0pt}map\ {\isacharcolon}{\kern0pt}{\isacharcolon}{\kern0pt}\ {\isachardoublequoteopen}{\isacharprime}{\kern0pt}e\isactrlsub {\isadigit{1}}\ {\isasymRightarrow}\ {\isacharprime}{\kern0pt}e\isactrlsub {\isadigit{2}}{\isachardoublequoteclose}%
\end{isabelle}
  
  Note that, a graph morphism maps the graph structure \isa{{\isacharparenleft}{\kern0pt}{\isacharprime}{\kern0pt}v\isactrlsub {\isadigit{1}}{\isacharcomma}{\kern0pt}\ {\isacharprime}{\kern0pt}e\isactrlsub {\isadigit{1}}{\isacharcomma}{\kern0pt}\ {\isacharprime}{\kern0pt}c{\isacharcomma}{\kern0pt}\ {\isacharprime}{\kern0pt}d{\isacharparenright}{\kern0pt}\ pre{\isacharunderscore}{\kern0pt}graph} 
  to \isa{{\isacharparenleft}{\kern0pt}{\isacharprime}{\kern0pt}v\isactrlsub {\isadigit{2}}{\isacharcomma}{\kern0pt}\ {\isacharprime}{\kern0pt}e\isactrlsub {\isadigit{2}}{\isacharcomma}{\kern0pt}\ {\isacharprime}{\kern0pt}c{\isacharcomma}{\kern0pt}\ {\isacharprime}{\kern0pt}d{\isacharparenright}{\kern0pt}\ pre{\isacharunderscore}{\kern0pt}graph}, i.e., the node and edge types change.
  Again, a common notation \isa{\isactrlbsub {\isacharunderscore}{\kern0pt}\isactrlesub \isactrlsub V} and \isa{\isactrlbsub {\isacharunderscore}{\kern0pt}\isactrlesub \isactrlsub E} for a morphism record is introduced
  using the \isacommand{notation} keyword.
  
  The locale \isa{morphism} inherits properties from the \isa{graph} locale via the
  \emph{import} mechanism. With this, a morphism carries its domain (\isa{G{\isacharcolon}{\kern0pt}\ graph\ G}) and its codomain 
  (\isa{H{\isacharcolon}{\kern0pt}\ graph\ H}) and all properties (i.e., all \isa{graph}, specialized for the particular
  instance), are inherited. The \isa{pre{\isacharunderscore}{\kern0pt}morph} record type is used to introduce a locale parameter \isa{f},
  which contains the corresponding node and edge mappings.%
\end{isamarkuptext}\isamarkuptrue%
\begin{isamarkuptext}%
The morphism properties are enforced by the following axioms:

\begin{itemize}%
\item \emph{Range restriction}, \isa{\mbox{}\inferrule{\mbox{e\ {\isasymin}\ E\isactrlbsub G\isactrlesub }}{\mbox{\isactrlbsub f\isactrlesub \isactrlsub E\ e\ {\isasymin}\ E\isactrlbsub H\isactrlesub }}} and \isa{\mbox{}\inferrule{\mbox{v\ {\isasymin}\ V\isactrlbsub G\isactrlesub }}{\mbox{\isactrlbsub f\isactrlesub \isactrlsub V\ v\ {\isasymin}\ V\isactrlbsub H\isactrlesub }}}

\item \emph{Source and target preservation}, \isa{\mbox{}\inferrule{\mbox{e\ {\isasymin}\ E\isactrlbsub G\isactrlesub }}{\mbox{\isactrlbsub f\isactrlesub \isactrlsub V\ {\isacharparenleft}{\kern0pt}s\isactrlbsub G\isactrlesub \ e{\isacharparenright}{\kern0pt}\ {\isacharequal}{\kern0pt}\ s\isactrlbsub H\isactrlesub \ {\isacharparenleft}{\kern0pt}\isactrlbsub f\isactrlesub \isactrlsub E\ e{\isacharparenright}{\kern0pt}}}} and \isa{\mbox{}\inferrule{\mbox{e\ {\isasymin}\ E\isactrlbsub G\isactrlesub }}{\mbox{\isactrlbsub f\isactrlesub \isactrlsub V\ {\isacharparenleft}{\kern0pt}t\isactrlbsub G\isactrlesub \ e{\isacharparenright}{\kern0pt}\ {\isacharequal}{\kern0pt}\ t\isactrlbsub H\isactrlesub \ {\isacharparenleft}{\kern0pt}\isactrlbsub f\isactrlesub \isactrlsub E\ e{\isacharparenright}{\kern0pt}}}}

\item \emph{Label preservation}, \isa{\mbox{}\inferrule{\mbox{v\ {\isasymin}\ V\isactrlbsub G\isactrlesub }}{\mbox{l\isactrlbsub G\isactrlesub \ v\ {\isacharequal}{\kern0pt}\ l\isactrlbsub H\isactrlesub \ {\isacharparenleft}{\kern0pt}\isactrlbsub f\isactrlesub \isactrlsub V\ v{\isacharparenright}{\kern0pt}}}} and \isa{\mbox{}\inferrule{\mbox{e\ {\isasymin}\ E\isactrlbsub G\isactrlesub }}{\mbox{m\isactrlbsub G\isactrlesub \ e\ {\isacharequal}{\kern0pt}\ m\isactrlbsub H\isactrlesub \ {\isacharparenleft}{\kern0pt}\isactrlbsub f\isactrlesub \isactrlsub E\ e{\isacharparenright}{\kern0pt}}}}%
\end{itemize}%
\end{isamarkuptext}\isamarkuptrue%
\begin{isamarkuptext}%
The \isa{morphism} locale definition is given by:

\begin{isabelle}%
\isacommand{locale}\ morphism\ {\isacharequal}{\kern0pt}\isanewline
\ \ \ \ G{\isacharcolon}{\kern0pt}\ graph\ G\ {\isacharplus}{\kern0pt}\isanewline
\ \ \ \ H{\isacharcolon}{\kern0pt}\ graph\ H\ \isakeyword{for}\isanewline
\ \ \ \ \ \ G\ {\isacharcolon}{\kern0pt}{\isacharcolon}{\kern0pt}\ {\isachardoublequoteopen}{\isacharparenleft}{\kern0pt}{\isacharprime}{\kern0pt}v\isactrlsub {\isadigit{1}}{\isacharcomma}{\kern0pt}{\isacharprime}{\kern0pt}e\isactrlsub {\isadigit{1}}{\isacharcomma}{\kern0pt}{\isacharprime}{\kern0pt}l{\isacharcomma}{\kern0pt}{\isacharprime}{\kern0pt}m{\isacharparenright}{\kern0pt}\ pre{\isacharunderscore}{\kern0pt}graph{\isachardoublequoteclose}\ \isakeyword{and}\ \isanewline
\ \ \ \ \ \ H\ {\isacharcolon}{\kern0pt}{\isacharcolon}{\kern0pt}\ {\isachardoublequoteopen}{\isacharparenleft}{\kern0pt}{\isacharprime}{\kern0pt}v\isactrlsub {\isadigit{2}}{\isacharcomma}{\kern0pt}{\isacharprime}{\kern0pt}e\isactrlsub {\isadigit{2}}{\isacharcomma}{\kern0pt}{\isacharprime}{\kern0pt}l{\isacharcomma}{\kern0pt}{\isacharprime}{\kern0pt}m{\isacharparenright}{\kern0pt}\ pre{\isacharunderscore}{\kern0pt}graph{\isachardoublequoteclose}\ {\isacharplus}{\kern0pt}\isanewline
\ \ \ \ \isakeyword{fixes}\ \isanewline
\ \ \ \ \ \ f\ {\isacharcolon}{\kern0pt}{\isacharcolon}{\kern0pt}\ {\isachardoublequoteopen}{\isacharparenleft}{\kern0pt}{\isacharprime}{\kern0pt}v\isactrlsub {\isadigit{1}}{\isacharcomma}{\kern0pt}{\isacharprime}{\kern0pt}v\isactrlsub {\isadigit{2}}{\isacharcomma}{\kern0pt}{\isacharprime}{\kern0pt}e\isactrlsub {\isadigit{1}}{\isacharcomma}{\kern0pt}{\isacharprime}{\kern0pt}e\isactrlsub {\isadigit{2}}{\isacharparenright}{\kern0pt}\ pre{\isacharunderscore}{\kern0pt}morph{\isachardoublequoteclose}\ \ \isanewline
\ \ \ \ \isakeyword{assumes}\isanewline
\ \ \ \ \ \ morph{\isacharunderscore}{\kern0pt}edge{\isacharunderscore}{\kern0pt}range{\isacharcolon}{\kern0pt}\ {\isachardoublequoteopen}e\ {\isasymin}\ E\isactrlbsub G\isactrlesub \ {\isasymLongrightarrow}\ \isactrlbsub f\isactrlesub \isactrlsub E\ e\ {\isasymin}\ E\isactrlbsub H\isactrlesub {\isachardoublequoteclose}\ \isakeyword{and}\isanewline
\ \ \ \ \ \ morph{\isacharunderscore}{\kern0pt}node{\isacharunderscore}{\kern0pt}range{\isacharcolon}{\kern0pt}\ {\isachardoublequoteopen}v\ {\isasymin}\ V\isactrlbsub G\isactrlesub \ {\isasymLongrightarrow}\ \isactrlbsub f\isactrlesub \isactrlsub V\ v\ {\isasymin}\ V\isactrlbsub H\isactrlesub {\isachardoublequoteclose}\ \isakeyword{and}\isanewline
\ \ \ \ \ \ source{\isacharunderscore}{\kern0pt}preserve\ {\isacharcolon}{\kern0pt}\ {\isachardoublequoteopen}e\ {\isasymin}\ E\isactrlbsub G\isactrlesub \ {\isasymLongrightarrow}\ \isactrlbsub f\isactrlesub \isactrlsub V\ {\isacharparenleft}{\kern0pt}s\isactrlbsub G\isactrlesub \ e{\isacharparenright}{\kern0pt}\ {\isacharequal}{\kern0pt}\ s\isactrlbsub H\isactrlesub \ {\isacharparenleft}{\kern0pt}\isactrlbsub f\isactrlesub \isactrlsub E\ e{\isacharparenright}{\kern0pt}{\isachardoublequoteclose}\ \isakeyword{and}\isanewline
\ \ \ \ \ \ target{\isacharunderscore}{\kern0pt}preserve\ {\isacharcolon}{\kern0pt}\ {\isachardoublequoteopen}e\ {\isasymin}\ E\isactrlbsub G\isactrlesub \ {\isasymLongrightarrow}\ \isactrlbsub f\isactrlesub \isactrlsub V\ {\isacharparenleft}{\kern0pt}t\isactrlbsub G\isactrlesub \ e{\isacharparenright}{\kern0pt}\ {\isacharequal}{\kern0pt}\ t\isactrlbsub H\isactrlesub \ {\isacharparenleft}{\kern0pt}\isactrlbsub f\isactrlesub \isactrlsub E\ e{\isacharparenright}{\kern0pt}{\isachardoublequoteclose}\ \isakeyword{and}\isanewline
\ \ \ \ \ \ label{\isacharunderscore}{\kern0pt}preserve\ \ {\isacharcolon}{\kern0pt}\ {\isachardoublequoteopen}v\ {\isasymin}\ V\isactrlbsub G\isactrlesub \ {\isasymLongrightarrow}\ l\isactrlbsub G\isactrlesub \ v\ {\isacharequal}{\kern0pt}\ l\isactrlbsub H\isactrlesub \ {\isacharparenleft}{\kern0pt}\isactrlbsub f\isactrlesub \isactrlsub V\ v{\isacharparenright}{\kern0pt}{\isachardoublequoteclose}\ \isakeyword{and}\isanewline
\ \ \ \ \ \ mark{\isacharunderscore}{\kern0pt}preserve\ \ \ {\isacharcolon}{\kern0pt}\ {\isachardoublequoteopen}e\ {\isasymin}\ E\isactrlbsub G\isactrlesub \ {\isasymLongrightarrow}\ m\isactrlbsub G\isactrlesub \ e\ {\isacharequal}{\kern0pt}\ m\isactrlbsub H\isactrlesub \ {\isacharparenleft}{\kern0pt}\isactrlbsub f\isactrlesub \isactrlsub E\ e{\isacharparenright}{\kern0pt}{\isachardoublequoteclose}%
\end{isabelle}

  With this, we define the composition of graph morphisms (cf. Def.~\ref{def:morphcomp}) 
  including the infix notation \isa{{\isasymcirc}\isactrlsub {\isasymrightarrow}} as the pairwise compositions:
\begin{isabelle}%
\isacommand{definition}\ morph{\isacharunderscore}{\kern0pt}comp\ \isanewline
\ \ \ \ \ \ {\isacharcolon}{\kern0pt}{\isacharcolon}{\kern0pt}\ {\isachardoublequoteopen}{\isacharparenleft}{\kern0pt}{\isacharprime}{\kern0pt}v\isactrlsub {\isadigit{2}}{\isacharcomma}{\kern0pt}{\isacharprime}{\kern0pt}v\isactrlsub {\isadigit{3}}{\isacharcomma}{\kern0pt}{\isacharprime}{\kern0pt}e\isactrlsub {\isadigit{2}}{\isacharcomma}{\kern0pt}{\isacharprime}{\kern0pt}e\isactrlsub {\isadigit{3}}{\isacharparenright}{\kern0pt}\ pre{\isacharunderscore}{\kern0pt}morph\ \ {\isasymRightarrow}\ \ {\isacharparenleft}{\kern0pt}{\isacharprime}{\kern0pt}v\isactrlsub {\isadigit{1}}{\isacharcomma}{\kern0pt}{\isacharprime}{\kern0pt}v\isactrlsub {\isadigit{2}}{\isacharcomma}{\kern0pt}{\isacharprime}{\kern0pt}e\isactrlsub {\isadigit{1}}{\isacharcomma}{\kern0pt}{\isacharprime}{\kern0pt}e\isactrlsub {\isadigit{2}}{\isacharparenright}{\kern0pt}\ pre{\isacharunderscore}{\kern0pt}morph\ \ {\isasymRightarrow}{\isacharparenleft}{\kern0pt}{\isacharprime}{\kern0pt}v\isactrlsub {\isadigit{1}}{\isacharcomma}{\kern0pt}{\isacharprime}{\kern0pt}v\isactrlsub {\isadigit{3}}{\isacharcomma}{\kern0pt}{\isacharprime}{\kern0pt}e\isactrlsub {\isadigit{1}}{\isacharcomma}{\kern0pt}{\isacharprime}{\kern0pt}e\isactrlsub {\isadigit{3}}{\isacharparenright}{\kern0pt}\ pre{\isacharunderscore}{\kern0pt}morph{\isachardoublequoteclose}\ {\isacharparenleft}{\kern0pt}\isakeyword{infixl}\ {\isachardoublequoteopen}{\isasymcirc}\isactrlsub {\isasymrightarrow}{\isachardoublequoteclose}\ {\isadigit{5}}{\isadigit{5}}{\isacharparenright}{\kern0pt}\ \isakeyword{where}\isanewline
\ \ \ \ {\isachardoublequoteopen}g\ {\isasymcirc}\isactrlsub {\isasymrightarrow}\ f\ {\isacharequal}{\kern0pt}\ {\isasymlparr}node{\isacharunderscore}{\kern0pt}map\ {\isacharequal}{\kern0pt}\ \isactrlbsub g\isactrlesub \isactrlsub V\ {\isasymcirc}\ \isactrlbsub f\isactrlesub \isactrlsub V{\isacharcomma}{\kern0pt}\ edge{\isacharunderscore}{\kern0pt}map\ {\isacharequal}{\kern0pt}\ \isactrlbsub g\isactrlesub \isactrlsub E\ {\isasymcirc}\ \isactrlbsub f\isactrlesub \isactrlsub E{\isasymrparr}{\isachardoublequoteclose}%
\end{isabelle}

  The proposition, from \isa{morphism\ G\ H\ f} and \isa{morphism\ H\ K\ g}
  we can conclude \isa{morphism\ G\ K\ {\isacharparenleft}{\kern0pt}g\ {\isasymcirc}\isactrlsub {\isasymrightarrow}\ f{\isacharparenright}{\kern0pt}} is expressed using the \emph{Isar} language as follows:%
\end{isamarkuptext}\isamarkuptrue%
\isacommand{lemma}\isamarkupfalse%
\isanewline
\ \ \isakeyword{assumes}\ f{\isacharcolon}{\kern0pt}\ {\isacartoucheopen}morphism\ G\ H\ f{\isacartoucheclose}\ \isakeyword{and}\ g{\isacharcolon}{\kern0pt}\ {\isacartoucheopen}morphism\ H\ K\ g{\isacartoucheclose}\isanewline
\ \ \isakeyword{shows}\ {\isacartoucheopen}morphism\ G\ K\ {\isacharparenleft}{\kern0pt}g\ {\isasymcirc}\isactrlsub {\isasymrightarrow}\ f{\isacharparenright}{\kern0pt}{\isacartoucheclose}%
\begin{isamarkuptext}%
Each premise, indicated by the \isa{assumes} keyword, is (optionally) associated with a name \isa{f} 
  and \isa{g}, respectively. The conclusion is indicated by the \isa{shows} statement. 
  We enter the proof by the \isacommand{proof} command followed by an optional proof method. 
  In this particular case, we use the \isa{intro{\isacharunderscore}{\kern0pt}locales} method, which applies the introduction rules
  of locales.%
\end{isamarkuptext}\isamarkuptrue%
\isadelimproof
\endisadelimproof
\isatagproof
\isacommand{proof}\isamarkupfalse%
\ intro{\isacharunderscore}{\kern0pt}locales%
\begin{isamarkuptext}%
Isabelle generates the following subgoals to discharge the lemma: %
\begin{isabelle}%
\ {\isadigit{1}}{\isachardot}{\kern0pt}\ graph\ G\isanewline
\ {\isadigit{2}}{\isachardot}{\kern0pt}\ graph\ K\isanewline
\ {\isadigit{3}}{\isachardot}{\kern0pt}\ morphism{\isacharunderscore}{\kern0pt}axioms\ G\ K\ {\isacharparenleft}{\kern0pt}g\ {\isasymcirc}\isactrlsub {\isasymrightarrow}\ f{\isacharparenright}{\kern0pt}%
\end{isabelle}
  The first two subgoals follow directly from the locale definition of \isa{morphisms}. The proof of
  \isa{graph\ G} is give by supplying the corresponding fact:%
\end{isamarkuptext}\isamarkuptrue%
\ \ \isacommand{show}\isamarkupfalse%
\ {\isacartoucheopen}graph\ G{\isacartoucheclose}\ \isacommand{by}\isamarkupfalse%
\ {\isacharparenleft}{\kern0pt}fact\ morphism{\isachardot}{\kern0pt}axioms{\isacharbrackleft}{\kern0pt}OF\ f{\isacharbrackright}{\kern0pt}{\isacharparenright}{\kern0pt}%
\begin{isamarkuptext}%
The \isa{morphism{\isachardot}{\kern0pt}axioms} definition is generated by the locale approach covering the stated
  locale assumptions. The \isa{OF} command is used to apply one theorem to another.
  The subgoal \isa{graph\ K} follows analogously.
  To prove the morphism axioms (\isa{morphism{\isacharunderscore}{\kern0pt}axioms}), the \isa{morphism{\isacharunderscore}{\kern0pt}axioms{\isachardot}{\kern0pt}intro} introduction 
  rule (generated by Isabelle) is used. Its definition is as follows:
  \begin{center}
    \isa{\mbox{}\inferrule{\mbox{{\isasymAnd}e{\isachardot}{\kern0pt}\ \mbox{}\inferrule{\mbox{e\ {\isasymin}\ E\isactrlbsub G\isactrlesub }}{\mbox{\isactrlbsub g\ {\isasymcirc}\isactrlsub {\isasymrightarrow}\ f\isactrlesub \isactrlsub E\ e\ {\isasymin}\ E\isactrlbsub K\isactrlesub }}}\\\ \mbox{{\isasymAnd}v{\isachardot}{\kern0pt}\ \mbox{}\inferrule{\mbox{v\ {\isasymin}\ V\isactrlbsub G\isactrlesub }}{\mbox{\isactrlbsub g\ {\isasymcirc}\isactrlsub {\isasymrightarrow}\ f\isactrlesub \isactrlsub V\ v\ {\isasymin}\ V\isactrlbsub K\isactrlesub }}}\\\ \mbox{{\isasymAnd}e{\isachardot}{\kern0pt}\ \mbox{}\inferrule{\mbox{e\ {\isasymin}\ E\isactrlbsub G\isactrlesub }}{\mbox{\isactrlbsub g\ {\isasymcirc}\isactrlsub {\isasymrightarrow}\ f\isactrlesub \isactrlsub V\ {\isacharparenleft}{\kern0pt}s\isactrlbsub G\isactrlesub \ e{\isacharparenright}{\kern0pt}\ {\isacharequal}{\kern0pt}\ s\isactrlbsub K\isactrlesub \ {\isacharparenleft}{\kern0pt}\isactrlbsub g\ {\isasymcirc}\isactrlsub {\isasymrightarrow}\ f\isactrlesub \isactrlsub E\ e{\isacharparenright}{\kern0pt}}}}\\\ \mbox{{\isasymAnd}e{\isachardot}{\kern0pt}\ \mbox{}\inferrule{\mbox{e\ {\isasymin}\ E\isactrlbsub G\isactrlesub }}{\mbox{\isactrlbsub g\ {\isasymcirc}\isactrlsub {\isasymrightarrow}\ f\isactrlesub \isactrlsub V\ {\isacharparenleft}{\kern0pt}t\isactrlbsub G\isactrlesub \ e{\isacharparenright}{\kern0pt}\ {\isacharequal}{\kern0pt}\ t\isactrlbsub K\isactrlesub \ {\isacharparenleft}{\kern0pt}\isactrlbsub g\ {\isasymcirc}\isactrlsub {\isasymrightarrow}\ f\isactrlesub \isactrlsub E\ e{\isacharparenright}{\kern0pt}}}}\\\ \mbox{{\isasymAnd}v{\isachardot}{\kern0pt}\ \mbox{}\inferrule{\mbox{v\ {\isasymin}\ V\isactrlbsub G\isactrlesub }}{\mbox{l\isactrlbsub G\isactrlesub \ v\ {\isacharequal}{\kern0pt}\ l\isactrlbsub K\isactrlesub \ {\isacharparenleft}{\kern0pt}\isactrlbsub g\ {\isasymcirc}\isactrlsub {\isasymrightarrow}\ f\isactrlesub \isactrlsub V\ v{\isacharparenright}{\kern0pt}}}}\\\ \mbox{{\isasymAnd}e{\isachardot}{\kern0pt}\ \mbox{}\inferrule{\mbox{e\ {\isasymin}\ E\isactrlbsub G\isactrlesub }}{\mbox{m\isactrlbsub G\isactrlesub \ e\ {\isacharequal}{\kern0pt}\ m\isactrlbsub K\isactrlesub \ {\isacharparenleft}{\kern0pt}\isactrlbsub g\ {\isasymcirc}\isactrlsub {\isasymrightarrow}\ f\isactrlesub \isactrlsub E\ e{\isacharparenright}{\kern0pt}}}}}{\mbox{morphism{\isacharunderscore}{\kern0pt}axioms\ G\ K\ {\isacharparenleft}{\kern0pt}g\ {\isasymcirc}\isactrlsub {\isasymrightarrow}\ f{\isacharparenright}{\kern0pt}}}}
  \end{center}
  Both, the constant (\isa{morphism{\isacharunderscore}{\kern0pt}axioms}) rule and the introduction rule, are generated by the \isacommand{locale} mechanism.%
\end{isamarkuptext}\isamarkuptrue%
\ \ \isacommand{show}\isamarkupfalse%
\ {\isacartoucheopen}morphism{\isacharunderscore}{\kern0pt}axioms\ G\ K\ {\isacharparenleft}{\kern0pt}g\ {\isasymcirc}\isactrlsub {\isasymrightarrow}\ f{\isacharparenright}{\kern0pt}{\isacartoucheclose}\ \isanewline
\ \ \isacommand{proof}\isamarkupfalse%
\begin{isamarkuptext}%
The \isacommand{proof} command, without an explicit proof method will use the \isa{standard} method.
  This method uses a heuristic to apply certain proof rules. In this 
  particular case, the introduction rule is used which results the following subgoals:    
\begin{isabelle}%
\ {\isadigit{1}}{\isachardot}{\kern0pt}\ {\isasymAnd}e{\isachardot}{\kern0pt}\ e\ {\isasymin}\ E\isactrlbsub G\isactrlesub \ {\isasymLongrightarrow}\ \isactrlbsub g\ {\isasymcirc}\isactrlsub {\isasymrightarrow}\ f\isactrlesub \isactrlsub E\ e\ {\isasymin}\ E\isactrlbsub K\isactrlesub \isanewline
\ {\isadigit{2}}{\isachardot}{\kern0pt}\ {\isasymAnd}v{\isachardot}{\kern0pt}\ v\ {\isasymin}\ V\isactrlbsub G\isactrlesub \ {\isasymLongrightarrow}\ \isactrlbsub g\ {\isasymcirc}\isactrlsub {\isasymrightarrow}\ f\isactrlesub \isactrlsub V\ v\ {\isasymin}\ V\isactrlbsub K\isactrlesub \isanewline
\ {\isadigit{3}}{\isachardot}{\kern0pt}\ {\isasymAnd}e{\isachardot}{\kern0pt}\ e\ {\isasymin}\ E\isactrlbsub G\isactrlesub \ {\isasymLongrightarrow}\ \isactrlbsub g\ {\isasymcirc}\isactrlsub {\isasymrightarrow}\ f\isactrlesub \isactrlsub V\ {\isacharparenleft}{\kern0pt}s\isactrlbsub G\isactrlesub \ e{\isacharparenright}{\kern0pt}\ {\isacharequal}{\kern0pt}\ s\isactrlbsub K\isactrlesub \ {\isacharparenleft}{\kern0pt}\isactrlbsub g\ {\isasymcirc}\isactrlsub {\isasymrightarrow}\ f\isactrlesub \isactrlsub E\ e{\isacharparenright}{\kern0pt}\isanewline
\ {\isadigit{4}}{\isachardot}{\kern0pt}\ {\isasymAnd}e{\isachardot}{\kern0pt}\ e\ {\isasymin}\ E\isactrlbsub G\isactrlesub \ {\isasymLongrightarrow}\ \isactrlbsub g\ {\isasymcirc}\isactrlsub {\isasymrightarrow}\ f\isactrlesub \isactrlsub V\ {\isacharparenleft}{\kern0pt}t\isactrlbsub G\isactrlesub \ e{\isacharparenright}{\kern0pt}\ {\isacharequal}{\kern0pt}\ t\isactrlbsub K\isactrlesub \ {\isacharparenleft}{\kern0pt}\isactrlbsub g\ {\isasymcirc}\isactrlsub {\isasymrightarrow}\ f\isactrlesub \isactrlsub E\ e{\isacharparenright}{\kern0pt}\isanewline
\ {\isadigit{5}}{\isachardot}{\kern0pt}\ {\isasymAnd}v{\isachardot}{\kern0pt}\ v\ {\isasymin}\ V\isactrlbsub G\isactrlesub \ {\isasymLongrightarrow}\ l\isactrlbsub G\isactrlesub \ v\ {\isacharequal}{\kern0pt}\ l\isactrlbsub K\isactrlesub \ {\isacharparenleft}{\kern0pt}\isactrlbsub g\ {\isasymcirc}\isactrlsub {\isasymrightarrow}\ f\isactrlesub \isactrlsub V\ v{\isacharparenright}{\kern0pt}\isanewline
\ {\isadigit{6}}{\isachardot}{\kern0pt}\ {\isasymAnd}e{\isachardot}{\kern0pt}\ e\ {\isasymin}\ E\isactrlbsub G\isactrlesub \ {\isasymLongrightarrow}\ m\isactrlbsub G\isactrlesub \ e\ {\isacharequal}{\kern0pt}\ m\isactrlbsub K\isactrlesub \ {\isacharparenleft}{\kern0pt}\isactrlbsub g\ {\isasymcirc}\isactrlsub {\isasymrightarrow}\ f\isactrlesub \isactrlsub E\ e{\isacharparenright}{\kern0pt}%
\end{isabelle}

  Exemplary, we show that the composition \isa{g\ {\isasymcirc}\isactrlsub {\isasymrightarrow}\ f} maps an edge from \isa{G} to an edge from \isa{K}.
  This subgoal arises from the \isa{morph{\isacharunderscore}{\kern0pt}edge{\isacharunderscore}{\kern0pt}range} axiom.%
\end{isamarkuptext}\isamarkuptrue%
\ \ \ \ \isacommand{show}\isamarkupfalse%
\ {\isacartoucheopen}\isactrlbsub g\ {\isasymcirc}\isactrlsub {\isasymrightarrow}\ f\isactrlesub \isactrlsub E\ e\ {\isasymin}\ E\isactrlbsub K\isactrlesub {\isacartoucheclose}\ \isakeyword{if}\ {\isacartoucheopen}e\ {\isasymin}\ E\isactrlbsub G\isactrlesub {\isacartoucheclose}\ \isakeyword{for}\ e\isanewline
\ \ \ \ \ \ \isacommand{by}\isamarkupfalse%
\ {\isacharparenleft}{\kern0pt}simp\ add{\isacharcolon}{\kern0pt}\ morph{\isacharunderscore}{\kern0pt}comp{\isacharunderscore}{\kern0pt}def\ morphism{\isachardot}{\kern0pt}morph{\isacharunderscore}{\kern0pt}edge{\isacharunderscore}{\kern0pt}range{\isacharbrackleft}{\kern0pt}OF\ g{\isacharbrackright}{\kern0pt}\ morphism{\isachardot}{\kern0pt}morph{\isacharunderscore}{\kern0pt}edge{\isacharunderscore}{\kern0pt}range{\isacharbrackleft}{\kern0pt}OF\ f{\isacharbrackright}{\kern0pt}\ that{\isacharparenright}{\kern0pt}%
\begin{isamarkuptext}%
To prove this goal, we unfold the definition of \isa{{\isacharparenleft}{\kern0pt}{\isasymcirc}\isactrlsub {\isasymrightarrow}{\isacharparenright}{\kern0pt}} by telling the simplifier
  to consider the morphism composition definition (\isa{morph{\isacharunderscore}{\kern0pt}comp{\isacharunderscore}{\kern0pt}def}) fact. With the fact that both,
  the \isa{morph{\isacharunderscore}{\kern0pt}edge{\isacharunderscore}{\kern0pt}range} axiom hold for \isa{g} and \isa{f}, and built-in facts on function composition,
  the simplifier is able to discharge the goal.

  Proving that the composition preserves the sources follows similarly. We unfold the
  composition definition and supply the \isa{morph{\isacharunderscore}{\kern0pt}edge{\isacharunderscore}{\kern0pt}range} and \isa{source{\isacharunderscore}{\kern0pt}preserve}, specialized for 
  each morphism \isa{f} and \isa{g} to the simplifier:%
\end{isamarkuptext}\isamarkuptrue%
\ \ \ \ \isacommand{show}\isamarkupfalse%
\ {\isacartoucheopen}\isactrlbsub g\ {\isasymcirc}\isactrlsub {\isasymrightarrow}\ f\isactrlesub \isactrlsub V\ {\isacharparenleft}{\kern0pt}s\isactrlbsub G\isactrlesub \ e{\isacharparenright}{\kern0pt}\ {\isacharequal}{\kern0pt}\ s\isactrlbsub K\isactrlesub \ {\isacharparenleft}{\kern0pt}\isactrlbsub g\ {\isasymcirc}\isactrlsub {\isasymrightarrow}\ f\isactrlesub \isactrlsub E\ e{\isacharparenright}{\kern0pt}{\isacartoucheclose}\ \isakeyword{if}\ {\isacartoucheopen}e\ {\isasymin}\ E\isactrlbsub G\isactrlesub {\isacartoucheclose}\ \isakeyword{for}\ e\isanewline
\ \ \ \ \ \ \isacommand{by}\isamarkupfalse%
\ {\isacharparenleft}{\kern0pt}simp\ add{\isacharcolon}{\kern0pt}\ morph{\isacharunderscore}{\kern0pt}comp{\isacharunderscore}{\kern0pt}def\ \isanewline
\ \ \ \ \ \ \ \ \ \ \ \ morphism{\isachardot}{\kern0pt}morph{\isacharunderscore}{\kern0pt}edge{\isacharunderscore}{\kern0pt}range{\isacharbrackleft}{\kern0pt}OF\ f{\isacharbrackright}{\kern0pt}\ \isanewline
\ \ \ \ \ \ \ \ \ \ \ \ morphism{\isachardot}{\kern0pt}morph{\isacharunderscore}{\kern0pt}edge{\isacharunderscore}{\kern0pt}range{\isacharbrackleft}{\kern0pt}OF\ g{\isacharbrackright}{\kern0pt}\ \isanewline
\ \ \ \ \ \ \ \ \ \ \ \ morphism{\isachardot}{\kern0pt}source{\isacharunderscore}{\kern0pt}preserve{\isacharbrackleft}{\kern0pt}OF\ f{\isacharbrackright}{\kern0pt}\ \isanewline
\ \ \ \ \ \ \ \ \ \ \ \ morphism{\isachardot}{\kern0pt}source{\isacharunderscore}{\kern0pt}preserve{\isacharbrackleft}{\kern0pt}OF\ g{\isacharbrackright}{\kern0pt}\ that{\isacharparenright}{\kern0pt}%
\endisatagproof
{\isafoldproof}%
\isadelimproof
\endisadelimproof
\begin{isamarkuptext}%
Isabelle's simplifier is able to discharge the proof obligation with the supplied facts.
  The other subgoals follow analogously. 

  Based on the \isa{morphism} locale, we formalise injective graph morphisms
 (cf. Def.~\ref{def:special-morph}) in a separate locale as follows:
\begin{isabelle}%
\isacommand{locale}\ injective{\isacharunderscore}{\kern0pt}morphism\ {\isacharequal}{\kern0pt}\ morphism\ {\isacharplus}{\kern0pt}\isanewline
\ \ \ \ \isakeyword{assumes}\ \isanewline
\ \ \ \ \ \ inj{\isacharunderscore}{\kern0pt}nodes{\isacharcolon}{\kern0pt}\ {\isachardoublequoteopen}inj{\isacharunderscore}{\kern0pt}on\ \isactrlbsub f\isactrlesub \isactrlsub V\ V\isactrlbsub G\isactrlesub {\isachardoublequoteclose}\ \isakeyword{and}\isanewline
\ \ \ \ \ \ inj{\isacharunderscore}{\kern0pt}edges{\isacharcolon}{\kern0pt}\ {\isachardoublequoteopen}inj{\isacharunderscore}{\kern0pt}on\ \isactrlbsub f\isactrlesub \isactrlsub E\ E\isactrlbsub G\isactrlesub {\isachardoublequoteclose}%
\end{isabelle}

  The locale axioms (\isa{inj{\isacharunderscore}{\kern0pt}nodes} and \isa{inj{\isacharunderscore}{\kern0pt}edges}) are used to restrict the node
  and edge mappings to be injective over the corresponding domain using Isabelle's built-in 
  \isa{inj{\isacharunderscore}{\kern0pt}on} predicate.
  Furthermore, we define surjective and bijective morphisms in a similar way.%
\end{isamarkuptext}\isamarkuptrue%
\isadelimdocument
\endisadelimdocument
\isatagdocument
\isamarkupsubsection{Pushouts \label{sec:pushouts}%
}
\isamarkuptrue%
\endisatagdocument
{\isafolddocument}%
\isadelimdocument
\endisadelimdocument
\begin{isamarkuptext}%
This subsection formalises pushouts in Isabelle and proves their uniqueness up to isomorphism
  (cf. Theorem \ref{theorem:po-uniqueness}). 
  The pushout characterisation comprises four commuting morphisms ($A \to B$, $A \to C$, $B \to D$,
  and $C \to D$) which satisfy the universal property (cf. Def.~\ref{def:pushout}). The commuting 
  property is expressed using the node \isa{node{\isacharunderscore}{\kern0pt}commutativity} and edge \isa{edge{\isacharunderscore}{\kern0pt}commutativity}
  proposition and the composition of morphisms. Our formalisation is given by:

\begin{isabelle}%
\isacommand{locale}\ pushout{\isacharunderscore}{\kern0pt}diagram\ {\isacharequal}{\kern0pt}\isanewline
\ \ b{\isacharcolon}{\kern0pt}\ morphism\ A\ B\ b\ {\isacharplus}{\kern0pt}\isanewline
\ \ c{\isacharcolon}{\kern0pt}\ morphism\ A\ C\ c\ {\isacharplus}{\kern0pt}\isanewline
\ \ f{\isacharcolon}{\kern0pt}\ morphism\ B\ D\ f\ {\isacharplus}{\kern0pt}\isanewline
\ \ g{\isacharcolon}{\kern0pt}\ morphism\ C\ D\ g\ \isakeyword{for}\ A\ B\ C\ \isakeyword{and}\ D\ {\isacharcolon}{\kern0pt}{\isacharcolon}{\kern0pt}\ {\isacartoucheopen}{\isacharparenleft}{\kern0pt}{\isacharprime}{\kern0pt}g{\isacharcomma}{\kern0pt}\ {\isacharprime}{\kern0pt}h{\isacharcomma}{\kern0pt}\ {\isacharprime}{\kern0pt}k{\isacharcomma}{\kern0pt}\ {\isacharprime}{\kern0pt}l{\isacharparenright}{\kern0pt}\ pre{\isacharunderscore}{\kern0pt}graph{\isacartoucheclose}\ \isakeyword{and}\ b\ c\ f\ g\ {\isacharplus}{\kern0pt}\isanewline
\isakeyword{assumes}\isanewline
\ \ node{\isacharunderscore}{\kern0pt}commutativity{\isacharcolon}{\kern0pt}\ {\isacartoucheopen}v\ {\isasymin}\ V\isactrlbsub A\isactrlesub \ {\isasymLongrightarrow}\ \isactrlbsub f\ {\isasymcirc}\isactrlsub {\isasymrightarrow}\ b\isactrlesub \isactrlsub V\ v\ {\isacharequal}{\kern0pt}\ \isactrlbsub g\ {\isasymcirc}\isactrlsub {\isasymrightarrow}\ c\isactrlesub \isactrlsub V\ v{\isacartoucheclose}\ \isakeyword{and}\isanewline
\ \ edge{\isacharunderscore}{\kern0pt}commutativity{\isacharcolon}{\kern0pt}\ {\isacartoucheopen}e\ {\isasymin}\ E\isactrlbsub A\isactrlesub \ {\isasymLongrightarrow}\ \isactrlbsub f\ {\isasymcirc}\isactrlsub {\isasymrightarrow}\ b\isactrlesub \isactrlsub E\ e\ {\isacharequal}{\kern0pt}\ \isactrlbsub g\ {\isasymcirc}\isactrlsub {\isasymrightarrow}\ c\isactrlesub \isactrlsub E\ e{\isacartoucheclose}\ \isakeyword{and}\isanewline
\ \ universal{\isacharunderscore}{\kern0pt}property{\isacharcolon}{\kern0pt}\ {\isacartoucheopen}{\isasymlbrakk}\isanewline
\ \ \ \ graph\ {\isacharparenleft}{\kern0pt}D{\isacharprime}{\kern0pt}\ {\isacharcolon}{\kern0pt}{\isacharcolon}{\kern0pt}\ {\isacharparenleft}{\kern0pt}{\isacharprime}{\kern0pt}g{\isacharcomma}{\kern0pt}{\isacharprime}{\kern0pt}h{\isacharcomma}{\kern0pt}{\isacharprime}{\kern0pt}k{\isacharcomma}{\kern0pt}{\isacharprime}{\kern0pt}l{\isacharparenright}{\kern0pt}\ pre{\isacharunderscore}{\kern0pt}graph{\isacharparenright}{\kern0pt}{\isacharsemicolon}{\kern0pt}\ \isanewline
\ \ \ \ morphism\ B\ D{\isacharprime}{\kern0pt}\ x{\isacharsemicolon}{\kern0pt}\ \isanewline
\ \ \ \ morphism\ C\ D{\isacharprime}{\kern0pt}\ y{\isacharsemicolon}{\kern0pt}\isanewline
\ \ \ \ \ {\isasymforall}v\ {\isasymin}\ V\isactrlbsub A\isactrlesub {\isachardot}{\kern0pt}\ \isactrlbsub x\ {\isasymcirc}\isactrlsub {\isasymrightarrow}\ b\isactrlesub \isactrlsub V\ v\ {\isacharequal}{\kern0pt}\ \isactrlbsub y\ {\isasymcirc}\isactrlsub {\isasymrightarrow}\ c\isactrlesub \isactrlsub V\ v{\isacharsemicolon}{\kern0pt}\isanewline
\ \ \ \ \ {\isasymforall}e\ {\isasymin}\ E\isactrlbsub A\isactrlesub {\isachardot}{\kern0pt}\ \isactrlbsub x\ {\isasymcirc}\isactrlsub {\isasymrightarrow}\ b\isactrlesub \isactrlsub E\ e\ {\isacharequal}{\kern0pt}\ \isactrlbsub y\ {\isasymcirc}\isactrlsub {\isasymrightarrow}\ c\isactrlesub \isactrlsub E\ e{\isasymrbrakk}\ \isanewline
\ \ \ \ \ \ {\isasymLongrightarrow}\ Ex{\isadigit{1}}M\ {\isacharparenleft}{\kern0pt}{\isasymlambda}u{\isachardot}{\kern0pt}\ morphism\ D\ D{\isacharprime}{\kern0pt}\ u\ {\isasymand}\isanewline
\ \ \ \ \ \ \ \ \ \ \ \ {\isacharparenleft}{\kern0pt}{\isasymforall}v\ {\isasymin}\ V\isactrlbsub B\isactrlesub {\isachardot}{\kern0pt}\ \isactrlbsub u\ {\isasymcirc}\isactrlsub {\isasymrightarrow}\ f\isactrlesub \isactrlsub V\ v\ {\isacharequal}{\kern0pt}\ \isactrlbsub x\isactrlesub \isactrlsub V\ v{\isacharparenright}{\kern0pt}\ {\isasymand}\isanewline
\ \ \ \ \ \ \ \ \ \ \ \ {\isacharparenleft}{\kern0pt}{\isasymforall}e\ {\isasymin}\ E\isactrlbsub B\isactrlesub {\isachardot}{\kern0pt}\ \isactrlbsub u\ {\isasymcirc}\isactrlsub {\isasymrightarrow}\ f\isactrlesub \isactrlsub E\ e\ {\isacharequal}{\kern0pt}\ \isactrlbsub x\isactrlesub \isactrlsub E\ e{\isacharparenright}{\kern0pt}\ {\isasymand}\isanewline
\ \ \ \ \ \ \ \ \ \ \ \ {\isacharparenleft}{\kern0pt}{\isasymforall}v\ {\isasymin}\ V\isactrlbsub C\isactrlesub {\isachardot}{\kern0pt}\ \isactrlbsub u\ {\isasymcirc}\isactrlsub {\isasymrightarrow}\ g\isactrlesub \isactrlsub V\ v\ {\isacharequal}{\kern0pt}\ \isactrlbsub y\isactrlesub \isactrlsub V\ v{\isacharparenright}{\kern0pt}\ {\isasymand}\isanewline
\ \ \ \ \ \ \ \ \ \ \ \ {\isacharparenleft}{\kern0pt}{\isasymforall}e\ {\isasymin}\ E\isactrlbsub C\isactrlesub {\isachardot}{\kern0pt}\ \isactrlbsub u\ {\isasymcirc}\isactrlsub {\isasymrightarrow}\ g\isactrlesub \isactrlsub E\ e\ {\isacharequal}{\kern0pt}\ \isactrlbsub y\isactrlesub \isactrlsub E\ e{\isacharparenright}{\kern0pt}{\isacharparenright}{\kern0pt}\isanewline
\ \ \ \ \ \ \ \ \ \ \ \ D{\isacartoucheclose}%
\end{isabelle}

  In Isabelle, unbound variables are implicitly bound using the (meta) universal quantifier. 
  For a given \isa{P}, the proposition \isa{P\ x} is interpreted as \isa{{\isasymAnd}x{\isachardot}{\kern0pt}\ P\ x}.
  Additionally, we restrict the universal quantified \isa{pre{\isacharunderscore}{\kern0pt}graph} record to match the 
  pushout object (graph \isa{D}) type. This prevents a warning generated by Isabelle's locale
  mechanism of newly introduced type-parameter. 
  We discuss this implication in Section~\ref{sec:conclusion}.

  Compared to our initial version of this paper \cite{Soeldner-Plump22a}, the usage of
  total functions increased the complexity of the \isa{pushout{\isacharunderscore}{\kern0pt}diagram} definition. The
  \isa{universal{\isacharunderscore}{\kern0pt}property}, stating the existence of a unique morphisms 
  (cf. Definition \ref{def:pushout}), cannot use the built-in unique existence  operator (\isa{Ex{\isadigit{1}}}).
  The proof of this operator would result in a goal to prove equality of the morphism \isa{u} from \isa{D} 
  to \isa{D{\isacharprime}{\kern0pt}}. As function equality (i.e. \isa{u\ {\isacharequal}{\kern0pt}\ u{\isacharprime}{\kern0pt}}) requires equality across the 
  universe of values of the domain (using the built-in simplification rule \isa{fun{\isacharunderscore}{\kern0pt}eq{\isacharunderscore}{\kern0pt}iff}, which
  is defined as \isa{{\isacharparenleft}{\kern0pt}u\ {\isacharequal}{\kern0pt}\ u{\isacharprime}{\kern0pt}{\isacharparenright}{\kern0pt}\ {\isacharequal}{\kern0pt}\ {\isacharparenleft}{\kern0pt}{\isasymforall}x{\isachardot}{\kern0pt}\ u\ x\ {\isacharequal}{\kern0pt}\ u{\isacharprime}{\kern0pt}\ x{\isacharparenright}{\kern0pt}}).
  In our formalisation, equality over the entire domain is not true. 
  We discuss this implication in Section \ref{sec:conclusion}.

  Our solution quantifies over the set of values (i.e., the set of nodes and edges of the source graph).
  Therefore, we introduce the abbreviation %
\begin{isabelle}%
Ex{\isadigit{1}}M\ {\isasymequiv}\isanewline
{\isasymlambda}P\ E{\isachardot}{\kern0pt}\ {\isasymexists}x{\isachardot}{\kern0pt}\ P\ x\ {\isasymand}\ {\isacharparenleft}{\kern0pt}{\isasymforall}y{\isachardot}{\kern0pt}\ P\ y\ {\isasymlongrightarrow}\ {\isacharparenleft}{\kern0pt}{\isasymforall}e{\isasymin}E\isactrlbsub E\isactrlesub {\isachardot}{\kern0pt}\ \isactrlbsub y\isactrlesub \isactrlsub E\ e\ {\isacharequal}{\kern0pt}\ \isactrlbsub x\isactrlesub \isactrlsub E\ e{\isacharparenright}{\kern0pt}\ {\isasymand}\ {\isacharparenleft}{\kern0pt}{\isasymforall}v{\isasymin}V\isactrlbsub E\isactrlesub {\isachardot}{\kern0pt}\ \isactrlbsub y\isactrlesub \isactrlsub V\ v\ {\isacharequal}{\kern0pt}\ \isactrlbsub x\isactrlesub \isactrlsub V\ v{\isacharparenright}{\kern0pt}{\isacharparenright}{\kern0pt}%
\end{isabelle} which is a lambda term capturing
  the predicate \isa{P} and the source graph \isa{E} to quantify over the corresponding sets.%
\end{isamarkuptext}\isamarkuptrue%
\begin{isamarkuptext}%
To prove the uniqueness of the pushout object, we assume the \isa{pushout{\isacharunderscore}{\kern0pt}diagram} 
  of $A \to B$, $A \to C$, $B \to D$, and $C\to D$, then the graph \isa{D{\isacharprime}{\kern0pt}} together with two 
  morphisms $f' \colon B \to D'$ and $g' \colon C \to D'$ is a pushout if and only if there
  exists a bijection \isa{u} between \isa{D} and \isa{D{\isacharprime}{\kern0pt}} such that the triangles 
  commute ($\forall x \in B. u \circ f x = f' x$ and $\forall x \in D. u \circ g x = g' x$).

  We formalise this theorem in Isabelle as follows:%
\end{isamarkuptext}\isamarkuptrue%
\isacommand{theorem}\isamarkupfalse%
\ \ uniqueness{\isacharunderscore}{\kern0pt}po{\isacharcolon}{\kern0pt}\isanewline
\ \ \isakeyword{fixes}\ D{\isacharprime}{\kern0pt}\ {\isacharcolon}{\kern0pt}{\isacharcolon}{\kern0pt}\ {\isacartoucheopen}{\isacharparenleft}{\kern0pt}{\isacharprime}{\kern0pt}g{\isacharcomma}{\kern0pt}\ {\isacharprime}{\kern0pt}h{\isacharcomma}{\kern0pt}\ {\isacharprime}{\kern0pt}k{\isacharcomma}{\kern0pt}\ {\isacharprime}{\kern0pt}l{\isacharparenright}{\kern0pt}\ pre{\isacharunderscore}{\kern0pt}graph{\isacartoucheclose}\isanewline
\ \ \isakeyword{assumes}\ \isanewline
\ \ \ \ D{\isacharprime}{\kern0pt}{\isacharcolon}{\kern0pt}\ {\isacartoucheopen}graph\ D{\isacharprime}{\kern0pt}{\isacartoucheclose}\ \isakeyword{and}\ \isanewline
\ \ \ \ f{\isacharprime}{\kern0pt}{\isacharcolon}{\kern0pt}\ {\isacartoucheopen}morphism\ B\ D{\isacharprime}{\kern0pt}\ f{\isacharprime}{\kern0pt}{\isacartoucheclose}\ \isakeyword{and}\ \isanewline
\ \ \ \ g{\isacharprime}{\kern0pt}{\isacharcolon}{\kern0pt}\ {\isacartoucheopen}morphism\ C\ D{\isacharprime}{\kern0pt}\ g{\isacharprime}{\kern0pt}{\isacartoucheclose}\isanewline
\ \ \isakeyword{shows}\ {\isacartoucheopen}pushout{\isacharunderscore}{\kern0pt}diagram\ \ A\ B\ C\ D{\isacharprime}{\kern0pt}\ b\ c\ f{\isacharprime}{\kern0pt}\ g{\isacharprime}{\kern0pt}\ \isanewline
\ \ \ \ {\isasymlongleftrightarrow}\ {\isacharparenleft}{\kern0pt}{\isasymexists}u{\isachardot}{\kern0pt}\ bijective{\isacharunderscore}{\kern0pt}morphism\ D\ D{\isacharprime}{\kern0pt}\ u\ \isanewline
\ \ \ \ \ \ \ \ \ \ {\isasymand}\ {\isacharparenleft}{\kern0pt}{\isasymforall}v\ {\isasymin}\ V\isactrlbsub B\isactrlesub {\isachardot}{\kern0pt}\ \isactrlbsub u\ {\isasymcirc}\isactrlsub {\isasymrightarrow}\ f\isactrlesub \isactrlsub V\ v\ {\isacharequal}{\kern0pt}\ \isactrlbsub f{\isacharprime}{\kern0pt}\isactrlesub \isactrlsub V\ v{\isacharparenright}{\kern0pt}\ {\isasymand}\ {\isacharparenleft}{\kern0pt}{\isasymforall}e\ {\isasymin}\ E\isactrlbsub B\isactrlesub {\isachardot}{\kern0pt}\ \isactrlbsub u\ {\isasymcirc}\isactrlsub {\isasymrightarrow}\ f\isactrlesub \isactrlsub E\ e\ {\isacharequal}{\kern0pt}\ \isactrlbsub f{\isacharprime}{\kern0pt}\isactrlesub \isactrlsub E\ e{\isacharparenright}{\kern0pt}\isanewline
\ \ \ \ \ \ \ \ \ \ {\isasymand}\ {\isacharparenleft}{\kern0pt}{\isasymforall}v\ {\isasymin}\ V\isactrlbsub C\isactrlesub {\isachardot}{\kern0pt}\ \isactrlbsub u\ {\isasymcirc}\isactrlsub {\isasymrightarrow}\ g\isactrlesub \isactrlsub V\ v\ {\isacharequal}{\kern0pt}\ \isactrlbsub g{\isacharprime}{\kern0pt}\isactrlesub \isactrlsub V\ v{\isacharparenright}{\kern0pt}\ {\isasymand}\ {\isacharparenleft}{\kern0pt}{\isasymforall}e\ {\isasymin}\ E\isactrlbsub C\isactrlesub {\isachardot}{\kern0pt}\ \isactrlbsub u\ {\isasymcirc}\isactrlsub {\isasymrightarrow}\ g\isactrlesub \isactrlsub E\ e\ {\isacharequal}{\kern0pt}\ \isactrlbsub g{\isacharprime}{\kern0pt}\isactrlesub \isactrlsub E\ e{\isacharparenright}{\kern0pt}{\isacharparenright}{\kern0pt}{\isacartoucheclose}\isanewline
\isadelimproof
\endisadelimproof
\isatagproof
\isacommand{proof}\isamarkupfalse%
\begin{isamarkuptext}%
The (implicitly) applied proof method \isa{standard} applies the built-in introduction 
  rule \isa{iffI}, which is used for \emph{if and only if} proofs, resulting in the following two subgoals:
\begin{isabelle}%
\ {\isadigit{1}}{\isachardot}{\kern0pt}\ pushout{\isacharunderscore}{\kern0pt}diagram\ A\ B\ C\ D{\isacharprime}{\kern0pt}\ b\ c\ f{\isacharprime}{\kern0pt}\ g{\isacharprime}{\kern0pt}\ {\isasymLongrightarrow}\isanewline
\isaindent{\ {\isadigit{1}}{\isachardot}{\kern0pt}\ }{\isasymexists}u{\isachardot}{\kern0pt}\ bijective{\isacharunderscore}{\kern0pt}morphism\ D\ D{\isacharprime}{\kern0pt}\ u\ {\isasymand}\isanewline
\isaindent{\ {\isadigit{1}}{\isachardot}{\kern0pt}\ {\isasymexists}u{\isachardot}{\kern0pt}\ }{\isacharparenleft}{\kern0pt}{\isasymforall}v{\isasymin}V\isactrlbsub B\isactrlesub {\isachardot}{\kern0pt}\ \isactrlbsub u\ {\isasymcirc}\isactrlsub {\isasymrightarrow}\ f\isactrlesub \isactrlsub V\ v\ {\isacharequal}{\kern0pt}\ \isactrlbsub f{\isacharprime}{\kern0pt}\isactrlesub \isactrlsub V\ v{\isacharparenright}{\kern0pt}\ {\isasymand}\isanewline
\isaindent{\ {\isadigit{1}}{\isachardot}{\kern0pt}\ {\isasymexists}u{\isachardot}{\kern0pt}\ }{\isacharparenleft}{\kern0pt}{\isasymforall}e{\isasymin}E\isactrlbsub B\isactrlesub {\isachardot}{\kern0pt}\ \isactrlbsub u\ {\isasymcirc}\isactrlsub {\isasymrightarrow}\ f\isactrlesub \isactrlsub E\ e\ {\isacharequal}{\kern0pt}\ \isactrlbsub f{\isacharprime}{\kern0pt}\isactrlesub \isactrlsub E\ e{\isacharparenright}{\kern0pt}\ {\isasymand}\isanewline
\isaindent{\ {\isadigit{1}}{\isachardot}{\kern0pt}\ {\isasymexists}u{\isachardot}{\kern0pt}\ }{\isacharparenleft}{\kern0pt}{\isasymforall}v{\isasymin}V\isactrlbsub C\isactrlesub {\isachardot}{\kern0pt}\ \isactrlbsub u\ {\isasymcirc}\isactrlsub {\isasymrightarrow}\ g\isactrlesub \isactrlsub V\ v\ {\isacharequal}{\kern0pt}\ \isactrlbsub g{\isacharprime}{\kern0pt}\isactrlesub \isactrlsub V\ v{\isacharparenright}{\kern0pt}\ {\isasymand}\ {\isacharparenleft}{\kern0pt}{\isasymforall}e{\isasymin}E\isactrlbsub C\isactrlesub {\isachardot}{\kern0pt}\ \isactrlbsub u\ {\isasymcirc}\isactrlsub {\isasymrightarrow}\ g\isactrlesub \isactrlsub E\ e\ {\isacharequal}{\kern0pt}\ \isactrlbsub g{\isacharprime}{\kern0pt}\isactrlesub \isactrlsub E\ e{\isacharparenright}{\kern0pt}\isanewline
\ {\isadigit{2}}{\isachardot}{\kern0pt}\ {\isasymexists}u{\isachardot}{\kern0pt}\ bijective{\isacharunderscore}{\kern0pt}morphism\ D\ D{\isacharprime}{\kern0pt}\ u\ {\isasymand}\isanewline
\isaindent{\ {\isadigit{2}}{\isachardot}{\kern0pt}\ {\isasymexists}u{\isachardot}{\kern0pt}\ }{\isacharparenleft}{\kern0pt}{\isasymforall}v{\isasymin}V\isactrlbsub B\isactrlesub {\isachardot}{\kern0pt}\ \isactrlbsub u\ {\isasymcirc}\isactrlsub {\isasymrightarrow}\ f\isactrlesub \isactrlsub V\ v\ {\isacharequal}{\kern0pt}\ \isactrlbsub f{\isacharprime}{\kern0pt}\isactrlesub \isactrlsub V\ v{\isacharparenright}{\kern0pt}\ {\isasymand}\isanewline
\isaindent{\ {\isadigit{2}}{\isachardot}{\kern0pt}\ {\isasymexists}u{\isachardot}{\kern0pt}\ }{\isacharparenleft}{\kern0pt}{\isasymforall}e{\isasymin}E\isactrlbsub B\isactrlesub {\isachardot}{\kern0pt}\ \isactrlbsub u\ {\isasymcirc}\isactrlsub {\isasymrightarrow}\ f\isactrlesub \isactrlsub E\ e\ {\isacharequal}{\kern0pt}\ \isactrlbsub f{\isacharprime}{\kern0pt}\isactrlesub \isactrlsub E\ e{\isacharparenright}{\kern0pt}\ {\isasymand}\isanewline
\isaindent{\ {\isadigit{2}}{\isachardot}{\kern0pt}\ {\isasymexists}u{\isachardot}{\kern0pt}\ }{\isacharparenleft}{\kern0pt}{\isasymforall}v{\isasymin}V\isactrlbsub C\isactrlesub {\isachardot}{\kern0pt}\ \isactrlbsub u\ {\isasymcirc}\isactrlsub {\isasymrightarrow}\ g\isactrlesub \isactrlsub V\ v\ {\isacharequal}{\kern0pt}\ \isactrlbsub g{\isacharprime}{\kern0pt}\isactrlesub \isactrlsub V\ v{\isacharparenright}{\kern0pt}\ {\isasymand}\ {\isacharparenleft}{\kern0pt}{\isasymforall}e{\isasymin}E\isactrlbsub C\isactrlesub {\isachardot}{\kern0pt}\ \isactrlbsub u\ {\isasymcirc}\isactrlsub {\isasymrightarrow}\ g\isactrlesub \isactrlsub E\ e\ {\isacharequal}{\kern0pt}\ \isactrlbsub g{\isacharprime}{\kern0pt}\isactrlesub \isactrlsub E\ e{\isacharparenright}{\kern0pt}\ {\isasymLongrightarrow}\isanewline
\isaindent{\ {\isadigit{2}}{\isachardot}{\kern0pt}\ }pushout{\isacharunderscore}{\kern0pt}diagram\ A\ B\ C\ D{\isacharprime}{\kern0pt}\ b\ c\ f{\isacharprime}{\kern0pt}\ g{\isacharprime}{\kern0pt}%
\end{isabelle}
  The proofs of both subgoals are omitted, but can be found on the GitHub repository for this
  formalisation.
  In the upcoming section, we will describe our formalisation of the \emph{gluing} construction.%
\end{isamarkuptext}\isamarkuptrue%
\endisatagproof
{\isafoldproof}%
\isadelimproof
\endisadelimproof
\isadelimdocument
\endisadelimdocument
\isatagdocument
\isamarkupsubsection{Gluings are Pushouts \label{sec:gluing-pushouts}%
}
\isamarkuptrue%
\endisatagdocument
{\isafolddocument}%
\isadelimdocument
\endisadelimdocument
\begin{isamarkuptext}%
The locale \isa{gluing} is used as an environment with required preconditions, 
  i.e., two injective  morphisms as described in Lemma~\ref{lemma:gluing}.
  Our definition is as follows:
\begin{isabelle}%
\isacommand{locale}\ gluing\ {\isacharequal}{\kern0pt}\isanewline
\ \ d{\isacharcolon}{\kern0pt}\ injective{\isacharunderscore}{\kern0pt}morphism\ K\ D\ d\ {\isacharplus}{\kern0pt}\isanewline
\ \ r{\isacharcolon}{\kern0pt}\ injective{\isacharunderscore}{\kern0pt}morphism\ K\ R\ b\isanewline
\ \ \isakeyword{for}\ K\ D\ R\ d\ b%
\end{isabelle}

  Within the locales context, we first define the gluing construction (graph \isa{D} together with 
  injective morphisms \isa{h} and \isa{c}) as depicted in Fig.~\ref{fig:gluing}. 
  Subsequently, we prove pushout correspondence (cf. Theorem~\ref{theorem:gluing-po}) by 
  interpretation of the \isa{pushout{\isacharunderscore}{\kern0pt}diagram} locale. Note that, while the \isacommand{interpretation}
  command is used for temporal instantiations (limited to the current context block), the 
  \isacommand{sublocale} command is used to create persistent links between locales 
  (see \cite{isabelle-isar-ref}). We use this technique to prove Corollary~\ref{corollary:dd-po}.%
\end{isamarkuptext}\isamarkuptrue%
\begin{isamarkuptext}%
The gluing graph \isa{D} can be constructed using the disjoint union of the node (edge) set
  (cf. Lemma~\ref{lemma:gluing}). In our formalisation, we use the built-in sum type 
  \isa{{\isacharprime}{\kern0pt}a\ {\isacharplus}{\kern0pt}\ {\isacharprime}{\kern0pt}b} type, it comes with the two injective functions 
  \isa{Inl{\isacharcolon}{\kern0pt}{\isacharcolon}{\kern0pt}{\isacharprime}{\kern0pt}a\ {\isasymRightarrow}\ {\isacharprime}{\kern0pt}a\ {\isacharplus}{\kern0pt}\ {\isacharprime}{\kern0pt}b} and \isa{Inr{\isacharcolon}{\kern0pt}{\isacharcolon}{\kern0pt}{\isacharprime}{\kern0pt}b\ {\isasymRightarrow}\ {\isacharprime}{\kern0pt}a\ {\isacharplus}{\kern0pt}\ {\isacharprime}{\kern0pt}b} which correspond to
  the $i_A$ and $i_B$, see Section~\ref{sec:background}. The image of a set \isa{A} under a function \isa{f}
  is denoted by \isa{f\ {\isacharbackquote}{\kern0pt}\ A} (\emph{backtick} operator).

  The node set is constructed by using the image of the nodes of \isa{D} under the injection \isa{Inl}
  united with the image of the nodes \isa{R} without \isa{b\ {\isacharbackquote}{\kern0pt}\ K} under the injection \isa{Inr}:

\begin{isabelle}%
\isacommand{abbreviation}\ V\ \isakeyword{where}\ {\isacartoucheopen}V\ {\isasymequiv}\ Inl\ {\isacharbackquote}{\kern0pt}\ V\isactrlbsub D\isactrlesub \ {\isasymunion}\ Inr\ {\isacharbackquote}{\kern0pt}\ {\isacharparenleft}{\kern0pt}V\isactrlbsub R\isactrlesub \ {\isacharminus}{\kern0pt}\ \isactrlbsub b\isactrlesub \isactrlsub V\ {\isacharbackquote}{\kern0pt}\ V\isactrlbsub K\isactrlesub {\isacharparenright}{\kern0pt}{\isacartoucheclose}%
\end{isabelle}

  The edge set follows analogously.
  We use the \isacommand{fun} command to state the source (target) function, as well as the 
  labelling functions. It will try proving certain properties (e.g., termination) of the function 
  under investigation automatically. In case the automation fails, the user has to prove 
  these properties by hand. 

  In our cases, Isabelle is able to discharge all generated proof obligations automatically.

  For the definition of the source (target) function, the general idea is a case analysis on the 
  edge origin by pattern matching on \isa{Inl\ e} and \isa{Inr\ e} constructors. 
  The source function is defined by:

\begin{isabelle}%
\isacommand{fun}\ s\ \isakeyword{where}\isanewline
\ \ \ {\isachardoublequoteopen}s\ {\isacharparenleft}{\kern0pt}Inl\ e{\isacharparenright}{\kern0pt}\ {\isacharequal}{\kern0pt}\ Inl\ {\isacharparenleft}{\kern0pt}s\isactrlbsub D\isactrlesub \ e{\isacharparenright}{\kern0pt}{\isachardoublequoteclose}\isanewline
\ \ {\isacharbar}{\kern0pt}{\isachardoublequoteopen}s\ {\isacharparenleft}{\kern0pt}Inr\ e{\isacharparenright}{\kern0pt}\ {\isacharequal}{\kern0pt}\ {\isacharparenleft}{\kern0pt}if\ e\ {\isasymin}\ {\isacharparenleft}{\kern0pt}E\isactrlbsub R\isactrlesub \ {\isacharminus}{\kern0pt}\ \isactrlbsub b\isactrlesub \isactrlsub E\ {\isacharbackquote}{\kern0pt}\ E\isactrlbsub K\isactrlesub {\isacharparenright}{\kern0pt}\ {\isasymand}\ {\isacharparenleft}{\kern0pt}s\isactrlbsub R\isactrlesub \ e\ {\isasymin}\ \isactrlbsub b\isactrlesub \isactrlsub V\ {\isacharbackquote}{\kern0pt}\ V\isactrlbsub K\isactrlesub {\isacharparenright}{\kern0pt}\ \isanewline
\ \ \ \ then\ Inl\ {\isacharparenleft}{\kern0pt}\isactrlbsub d\isactrlesub \isactrlsub V\ {\isacharparenleft}{\kern0pt}{\isacharparenleft}{\kern0pt}inv{\isacharunderscore}{\kern0pt}into\ V\isactrlbsub K\isactrlesub \ \isactrlbsub b\isactrlesub \isactrlsub V{\isacharparenright}{\kern0pt}\ {\isacharparenleft}{\kern0pt}s\isactrlbsub R\isactrlesub \ e{\isacharparenright}{\kern0pt}{\isacharparenright}{\kern0pt}{\isacharparenright}{\kern0pt}\ else\ Inr\ {\isacharparenleft}{\kern0pt}s\isactrlbsub R\isactrlesub \ e{\isacharparenright}{\kern0pt}{\isacharparenright}{\kern0pt}{\isachardoublequoteclose}%
\end{isabelle}

  In the \isa{Inl\ e} case, by definition of the edge set, the edge \isa{e} belongs to the graph \isa{D}. 
  As a result, we use the source function of \isa{D}. Compared to Section~\ref{sec:background}, where
  we assume $i_A$ ($i_B$) are inclusions (to keep the section readable), we now have to be more
  explicit. In the \isa{Inr\ e} case, the inverse of \isa{b} is given by the built-in \isa{inv{\isacharunderscore}{\kern0pt}into} function.

  The target function is analogous. Both labelling functions (for nodes and edges) are defined by 
  case analysis of the origin. Exemplary, the node labelling function \isa{l} is given by:

\begin{isabelle}%
\isacommand{fun}\ l\ \isakeyword{where}\isanewline
\ \ \ \ {\isachardoublequoteopen}l\ {\isacharparenleft}{\kern0pt}Inl\ v{\isacharparenright}{\kern0pt}\ {\isacharequal}{\kern0pt}\ l\isactrlbsub D\isactrlesub \ v{\isachardoublequoteclose}\isanewline
\ \ {\isacharbar}{\kern0pt}\ {\isachardoublequoteopen}l\ {\isacharparenleft}{\kern0pt}Inr\ v{\isacharparenright}{\kern0pt}\ {\isacharequal}{\kern0pt}\ l\isactrlbsub R\isactrlesub \ v{\isachardoublequoteclose}%
\end{isabelle}

  The edge labelling function is defined analogously. We follow by proving that these elements indeed 
  form a graph (according to our locale \isa{graph}) by interpretation. 
  The proof is mechanic and follows by case splitting on the sum type.

  Further, we define the morphisms \(h \colon R \to H\) and \(c \colon D \to H\).
  The morphism \isa{h} is defined by a case distinction of the node (edge) parameter. If the node (edge)
  \isa{v} is in the set \isa{V\isactrlbsub R\isactrlesub \ {\isacharminus}{\kern0pt}\ \isactrlbsub b\isactrlesub \isactrlsub V\ {\isacharbackquote}{\kern0pt}\ V\isactrlbsub K\isactrlesub } we know it is a newly created node (edge) and therefore has
  to be lifted using the \isa{Inr} injection. Otherwise, the node (edge) is already in the graph \isa{D}.
  Due to the injective \isa{b} (compared to inclusion), we use the inverse of \isa{b} followed by \isa{d}.
  Finally, the node (edge) is lifted into the sum type by using the \isa{Inl} injection.

  Our definition is as follows:
\begin{isabelle}%
\isacommand{abbreviation}\ h\ \isakeyword{where}\ \isanewline
\ \ {\isacartoucheopen}h\ {\isasymequiv}\ {\isasymlparr}node{\isacharunderscore}{\kern0pt}map\ {\isacharequal}{\kern0pt}\ {\isasymlambda}v{\isachardot}{\kern0pt}\ if\ v\ {\isasymin}\ V\isactrlbsub R\isactrlesub \ {\isacharminus}{\kern0pt}\ \isactrlbsub b\isactrlesub \isactrlsub V\ {\isacharbackquote}{\kern0pt}\ V\isactrlbsub K\isactrlesub \ then\ Inr\ v\ else\ Inl\ {\isacharparenleft}{\kern0pt}\isactrlbsub d\isactrlesub \isactrlsub V\ {\isacharparenleft}{\kern0pt}{\isacharparenleft}{\kern0pt}inv{\isacharunderscore}{\kern0pt}into\ V\isactrlbsub K\isactrlesub \ \isactrlbsub b\isactrlesub \isactrlsub V{\isacharparenright}{\kern0pt}\ v{\isacharparenright}{\kern0pt}{\isacharparenright}{\kern0pt}{\isacharcomma}{\kern0pt}\isanewline
\ \ \ \ \ \ \ \ edge{\isacharunderscore}{\kern0pt}map\ {\isacharequal}{\kern0pt}\ {\isasymlambda}e{\isachardot}{\kern0pt}\ if\ e\ {\isasymin}\ E\isactrlbsub R\isactrlesub \ {\isacharminus}{\kern0pt}\ \isactrlbsub b\isactrlesub \isactrlsub E\ {\isacharbackquote}{\kern0pt}\ E\isactrlbsub K\isactrlesub \ then\ Inr\ e\ else\ Inl\ {\isacharparenleft}{\kern0pt}\isactrlbsub d\isactrlesub \isactrlsub E\ {\isacharparenleft}{\kern0pt}{\isacharparenleft}{\kern0pt}inv{\isacharunderscore}{\kern0pt}into\ E\isactrlbsub K\isactrlesub \ \isactrlbsub b\isactrlesub \isactrlsub E{\isacharparenright}{\kern0pt}\ e{\isacharparenright}{\kern0pt}{\isacharparenright}{\kern0pt}{\isasymrparr}{\isacartoucheclose}\isanewline
\ \ %
\end{isabelle}

  The morphism \isa{c} is defined by the node (edge) injection \isa{Inl}:
\begin{isabelle}%
\isacommand{abbreviation}\ c\ \isakeyword{where}\ {\isacartoucheopen}c\ {\isasymequiv}\ {\isasymlparr}node{\isacharunderscore}{\kern0pt}map\ {\isacharequal}{\kern0pt}\ Inl{\isacharcomma}{\kern0pt}\ edge{\isacharunderscore}{\kern0pt}map\ {\isacharequal}{\kern0pt}\ Inl{\isasymrparr}{\isacartoucheclose}%
\end{isabelle}

  Each time, we prove that our construction fulfills the \isa{injective{\isacharunderscore}{\kern0pt}morphism} axioms by 
  interpretation. The proofs can be found on GitHub (interpretation \isa{inj{\isacharunderscore}{\kern0pt}h} and \isa{inj{\isacharunderscore}{\kern0pt}c}).
  Finally, we are able to prove the pushout correspondence by instantiating 
  (using the \isacommand{sublocale} command) of the \isa{pushout{\isacharunderscore}{\kern0pt}diagram} locale with the 
  corresponding parameters:
\begin{isabelle}%
\isacommand{sublocale}\ po{\isacharcolon}{\kern0pt}\ pushout{\isacharunderscore}{\kern0pt}diagram\ K\ R\ D\ H\ b\ d\ h\ c%
\end{isabelle}

  Here, \isa{po} is used to refer to \isa{pushout{\isacharunderscore}{\kern0pt}diagram} instance. 
  The proof is around 170 lines of text.

  In the upcoming subsection, we describe our formalisation of the \emph{deletion} construction.%
\end{isamarkuptext}\isamarkuptrue%
\isadelimdocument
\endisadelimdocument
\isatagdocument
\isamarkupsubsection{Deletions are Pushouts \label{sec:deletion-pushouts}%
}
\isamarkuptrue%
\endisatagdocument
{\isafolddocument}%
\isadelimdocument
\endisadelimdocument
\begin{isamarkuptext}%
The \isa{deletion} locale is also used as an environment with the required preconditions
  as stated in Lemma~\ref{lemma:deletion}. 
  The dangling condition (cf. Def.~\ref{def:dang}) is expressed using separate rules for the 
  source (\isa{dang{\isacharunderscore}{\kern0pt}src}) and target (\isa{dang{\isacharunderscore}{\kern0pt}trg}) mapping as follows: 
  \begin{center}
    \isa{\mbox{}\inferrule{\mbox{e\ {\isasymin}\ E}}{\mbox{s\isactrlbsub G\isactrlesub \ e\ {\isasymnotin}\ \isactrlbsub g\isactrlesub \isactrlsub V\ {\isacharbackquote}{\kern0pt}\ {\isacharparenleft}{\kern0pt}V\isactrlbsub L\isactrlesub \ {\isacharminus}{\kern0pt}\ \isactrlbsub b{\isacharprime}{\kern0pt}\isactrlesub \isactrlsub V\ {\isacharbackquote}{\kern0pt}\ V\isactrlbsub K\isactrlesub {\isacharparenright}{\kern0pt}}}} \qquad
    \isa{\mbox{}\inferrule{\mbox{e\ {\isasymin}\ E}}{\mbox{t\isactrlbsub G\isactrlesub \ e\ {\isasymnotin}\ \isactrlbsub g\isactrlesub \isactrlsub V\ {\isacharbackquote}{\kern0pt}\ {\isacharparenleft}{\kern0pt}V\isactrlbsub L\isactrlesub \ {\isacharminus}{\kern0pt}\ \isactrlbsub b{\isacharprime}{\kern0pt}\isactrlesub \isactrlsub V\ {\isacharbackquote}{\kern0pt}\ V\isactrlbsub K\isactrlesub {\isacharparenright}{\kern0pt}}}}
  \end{center}

  These locale assumptions are introduced in the \emph{assumes} section as follows:
\begin{isabelle}%
\isacommand{locale}\ deletion\ {\isacharequal}{\kern0pt}\isanewline
\ \ \ \ g{\isacharcolon}{\kern0pt}\ injective{\isacharunderscore}{\kern0pt}morphism\ L\ G\ g\ {\isacharplus}{\kern0pt}\isanewline
\ \ \ \ l{\isacharcolon}{\kern0pt}\ injective{\isacharunderscore}{\kern0pt}morphism\ K\ L\ b{\isacharprime}{\kern0pt}\isanewline
\ \ \ \ \isakeyword{for}\ K\ G\ L\ g\ b{\isacharprime}{\kern0pt}\ {\isacharplus}{\kern0pt}\isanewline
\ \ \ \ \isakeyword{assumes}\isanewline
\ \ \ \ \ \ dang{\isacharunderscore}{\kern0pt}src{\isacharcolon}{\kern0pt}\ {\isacartoucheopen}e\ {\isasymin}\ E\isactrlbsub G\isactrlesub \ {\isacharminus}{\kern0pt}\ \isactrlbsub g\isactrlesub \isactrlsub E\ {\isacharbackquote}{\kern0pt}\ {\isacharparenleft}{\kern0pt}E\isactrlbsub L\isactrlesub \ {\isacharminus}{\kern0pt}\ \isactrlbsub b{\isacharprime}{\kern0pt}\isactrlesub \isactrlsub E\ {\isacharbackquote}{\kern0pt}\ E\isactrlbsub K\isactrlesub {\isacharparenright}{\kern0pt}\ {\isasymLongrightarrow}\ s\isactrlbsub G\isactrlesub \ e\ {\isasymnotin}\ \isactrlbsub g\isactrlesub \isactrlsub V\ {\isacharbackquote}{\kern0pt}\ {\isacharparenleft}{\kern0pt}V\isactrlbsub L\isactrlesub \ {\isacharminus}{\kern0pt}\ \isactrlbsub b{\isacharprime}{\kern0pt}\isactrlesub \isactrlsub V\ {\isacharbackquote}{\kern0pt}\ V\isactrlbsub K\isactrlesub {\isacharparenright}{\kern0pt}{\isacartoucheclose}\ \isakeyword{and}\isanewline
\ \ \ \ \ \ dang{\isacharunderscore}{\kern0pt}trg{\isacharcolon}{\kern0pt}\ {\isacartoucheopen}e\ {\isasymin}\ E\isactrlbsub G\isactrlesub \ {\isacharminus}{\kern0pt}\ \isactrlbsub g\isactrlesub \isactrlsub E\ {\isacharbackquote}{\kern0pt}\ {\isacharparenleft}{\kern0pt}E\isactrlbsub L\isactrlesub \ {\isacharminus}{\kern0pt}\ \isactrlbsub b{\isacharprime}{\kern0pt}\isactrlesub \isactrlsub E\ {\isacharbackquote}{\kern0pt}\ E\isactrlbsub K\isactrlesub {\isacharparenright}{\kern0pt}\ {\isasymLongrightarrow}\ t\isactrlbsub G\isactrlesub \ e\ {\isasymnotin}\ \isactrlbsub g\isactrlesub \isactrlsub V\ {\isacharbackquote}{\kern0pt}\ {\isacharparenleft}{\kern0pt}V\isactrlbsub L\isactrlesub \ {\isacharminus}{\kern0pt}\ \isactrlbsub b{\isacharprime}{\kern0pt}\isactrlesub \isactrlsub V\ {\isacharbackquote}{\kern0pt}\ V\isactrlbsub K\isactrlesub {\isacharparenright}{\kern0pt}{\isacartoucheclose}%
\end{isabelle}

  The construction of the graph \isa{D} removes all nodes (edges) from \isa{G} which belong to a
  subgraph \isa{L} (without the image of \isa{b} under \isa{K}) under the morphism \isa{g}. 
  The construction of the node set is given by:

\begin{isabelle}%
\isacommand{abbreviation}\ V\ \isakeyword{where}\ {\isacartoucheopen}V\ {\isasymequiv}\ V\isactrlbsub G\isactrlesub \ {\isacharminus}{\kern0pt}\ \isactrlbsub g\isactrlesub \isactrlsub V\ {\isacharbackquote}{\kern0pt}\ {\isacharparenleft}{\kern0pt}V\isactrlbsub L\isactrlesub \ {\isacharminus}{\kern0pt}\ \isactrlbsub b{\isacharprime}{\kern0pt}\isactrlesub \isactrlsub V\ {\isacharbackquote}{\kern0pt}\ V\isactrlbsub K\isactrlesub {\isacharparenright}{\kern0pt}{\isacartoucheclose}%
\end{isabelle}

  The edge set follows analogously. The graph \isa{D} (\isa{pre{\isacharunderscore}{\kern0pt}graph} record) is constructed by updating 
  graph \isa{G} with a restricted set of nodes and edges. We use the record update syntax 
  \isa{G{\isasymlparr}nodes{\isacharcolon}{\kern0pt}{\isacharequal}{\kern0pt}V{\isacharcomma}{\kern0pt}edges{\isacharcolon}{\kern0pt}{\isacharequal}{\kern0pt}E{\isasymrparr}}, which replaces the set of nodes (edges) by \isa{V} (\isa{E}). 
  The other graph elements (i.e., \isa{s}, \isa{t}, \isa{l}, and \isa{m}) do not have to be changed, as we quantify 
  over the corresponding node (edge) set and therefore limit our reasoning to the valid subset of
  nodes (edges).

  We define the  injective morphisms \(d \colon K \to D\) by the composition of \isa{g} after \isa{b{\isacharprime}{\kern0pt}}:
\begin{isabelle}%
\isacommand{abbreviation}\ d\ \isakeyword{where}\isanewline
\ \ {\isacartoucheopen}d\ {\isasymequiv}\ g\ {\isasymcirc}\isactrlsub {\isasymrightarrow}\ b{\isacharprime}{\kern0pt}{\isacartoucheclose}%
\end{isabelle}

  The \isa{injective{\isacharunderscore}{\kern0pt}morphism} interpretation (\isa{inj{\isacharunderscore}{\kern0pt}d}) can be found on GitHub.
  The proof relies on the specialised proposition of well-formed composition of graph 
  morphisms (\isa{wf{\isacharunderscore}{\kern0pt}morph{\isacharunderscore}{\kern0pt}comp}).

  The morphism \(c' \colon D \to G\) is an inclusion, which we define using the \isa{idM{\isacharcolon}{\kern0pt}{\isacharcolon}{\kern0pt}{\isacharparenleft}{\kern0pt}{\isacharprime}{\kern0pt}i{\isacharcomma}{\kern0pt}\ {\isacharprime}{\kern0pt}i{\isacharcomma}{\kern0pt}\ {\isacharprime}{\kern0pt}j{\isacharcomma}{\kern0pt}\ {\isacharprime}{\kern0pt}j{\isacharparenright}{\kern0pt}\ pre{\isacharunderscore}{\kern0pt}morph} as:
\begin{isabelle}%
\isacommand{abbreviation}\ c{\isacharprime}{\kern0pt}\ \isakeyword{where}\ {\isacartoucheopen}c{\isacharprime}{\kern0pt}\ {\isasymequiv}\ idM{\isacartoucheclose}%
\end{isabelle}

  The \isa{idM} record (defined in the \isa{DPO{\isacharminus}{\kern0pt}Formal{\isachardot}{\kern0pt}Morphism} theory) uses the built-in 
  identity function (\isa{id{\isacharcolon}{\kern0pt}{\isacharcolon}{\kern0pt}{\isacharprime}{\kern0pt}i\ {\isasymRightarrow}\ {\isacharprime}{\kern0pt}i}) for both, the node and edge
  mappings (i.e., \isa{{\isasymlparr}node{\isacharunderscore}{\kern0pt}map\ {\isacharequal}{\kern0pt}\ id{\isacharcomma}{\kern0pt}\ edge{\isacharunderscore}{\kern0pt}map\ {\isacharequal}{\kern0pt}\ id{\isasymrparr}}).\\
  Here, the \isa{injective{\isacharunderscore}{\kern0pt}morphism} interpretation can be solved by the simplifier using the
  \isa{simp{\isacharunderscore}{\kern0pt}all} proof method.

  Ultimately, we prove the pushout correspondence by interpretation of the \isa{pushout{\isacharunderscore}{\kern0pt}diagram}
  locale with the corresponding parameters:
\begin{isabelle}%
\isacommand{sublocale}\ po{\isacharcolon}{\kern0pt}\ pushout{\isacharunderscore}{\kern0pt}diagram\ K\ L\ D\ G\ b{\isacharprime}{\kern0pt}\ d\ g\ c{\isacharprime}{\kern0pt}%
\end{isabelle}

  The proof is around 350 lines of text and follows mainly by case analysis of the edge (node)
  origin. A concise proof, relying on the gluing construction of \isa{D} and \isa{L}, failed due to
  Isabelle's typechecker not accepting our definition. 
  Our \isa{gluing} construction results in a pushout object with the type signature 
  \isa{{\isacharparenleft}{\kern0pt}{\isacharprime}{\kern0pt}a\ {\isacharplus}{\kern0pt}\ {\isacharprime}{\kern0pt}e{\isacharcomma}{\kern0pt}\ {\isacharprime}{\kern0pt}b\ {\isacharplus}{\kern0pt}\ {\isacharprime}{\kern0pt}f{\isacharcomma}{\kern0pt}\ {\isacharprime}{\kern0pt}c{\isacharcomma}{\kern0pt}\ {\isacharprime}{\kern0pt}d{\isacharparenright}{\kern0pt}\ pre{\isacharunderscore}{\kern0pt}graph}. As a result, we cannot use the universal property
  for graph \isa{G} (with signature \isa{{\isacharparenleft}{\kern0pt}{\isacharprime}{\kern0pt}e{\isacharcomma}{\kern0pt}\ {\isacharprime}{\kern0pt}f{\isacharcomma}{\kern0pt}\ {\isacharprime}{\kern0pt}c{\isacharcomma}{\kern0pt}\ {\isacharprime}{\kern0pt}d{\isacharparenright}{\kern0pt}\ pre{\isacharunderscore}{\kern0pt}graph}) as these types mismatch. 
  We will comment in Section~\ref{sec:conclusion} on this.

  In the following subsection we introduce rules and the notation of direct derivations.%
\end{isamarkuptext}\isamarkuptrue%
\isadelimdocument
\endisadelimdocument
\isatagdocument
\isamarkupsubsection{Rules and Derivations \label{sec:rules-derivations}%
}
\isamarkuptrue%
\endisatagdocument
{\isafolddocument}%
\isadelimdocument
\endisadelimdocument
\begin{isamarkuptext}%
Our formalisation of rules (cf. Def.~\ref{def:rule}) relies on the inclusion \(K \to L\) and
  \(K \to R\) which we represent in the locale \isa{rule}:

\begin{isabelle}%
\isacommand{locale}\ rule\ {\isacharequal}{\kern0pt}\isanewline
\ \ \ \ k{\isacharcolon}{\kern0pt}\ inclusion{\isacharunderscore}{\kern0pt}morphism\ K\ L\ {\isacharplus}{\kern0pt}\isanewline
\ \ \ \ r{\isacharcolon}{\kern0pt}\ inclusion{\isacharunderscore}{\kern0pt}morphism\ K\ R\ \isanewline
\ \ \ \ \isakeyword{for}\ L\ K\ R\isanewline
\ \ \isakeyword{begin}%
\end{isabelle}

  As we have explicitly included the domain and codomain in our definition of morphisms, we 
  inherit all properties and Isabelle's locale mechanism allows this dense definition.

  Using the \isacommand{notation} we introduce the common syntactical representation \isa{L\ {\isasymleftarrow}\ K\ {\isasymrightarrow}\ R} for
  \isa{L}, \isa{K}, and \isa{R} of type \isa{{\isacharparenleft}{\kern0pt}{\isacharprime}{\kern0pt}v{\isacharcomma}{\kern0pt}\ {\isacharprime}{\kern0pt}e{\isacharcomma}{\kern0pt}\ {\isacharprime}{\kern0pt}l{\isacharcomma}{\kern0pt}\ {\isacharprime}{\kern0pt}m{\isacharparenright}{\kern0pt}\ pre{\isacharunderscore}{\kern0pt}graph} on the \isa{rule} predicate.

  A direct derivation (cf. Def.~\ref{def:direct-derivation}) is defined in terms of existing constructs;
  the \isa{direct{\isacharunderscore}{\kern0pt}derivation} locale imports from the \isa{rule} and passes the pushout 
  complement from the \isa{deletion} locale (\isa{d{\isachardot}{\kern0pt}H}) into the \isa{gluing} locale to 
  construct pushout object \isa{g{\isachardot}{\kern0pt}h}. We model this in Isabelle as follows:

\begin{isabelle}%
\isacommand{locale}\ direct{\isacharunderscore}{\kern0pt}derivation\ {\isacharequal}{\kern0pt}\isanewline
\ \ \ \ r{\isacharcolon}{\kern0pt}\ rule\ L\ K\ R\ {\isacharplus}{\kern0pt}\isanewline
\ \ \ \ d{\isacharcolon}{\kern0pt}\ deletion\ K\ G\ L\ g\ idM\ {\isacharplus}{\kern0pt}\isanewline
\ \ \ \ g{\isacharcolon}{\kern0pt}\ gluing\ K\ d{\isachardot}{\kern0pt}D\ R\ g\ idM\ \isakeyword{for}\ G\ L\ K\ R\ g\isanewline
\ \ %
\end{isabelle}
  
  Within the \isa{direct{\isacharunderscore}{\kern0pt}derivation} context, we prove that given a direct derivation \(G \Rightarrow_{r,g} M\),
  squares (1) and (2) in Figure~\ref{fig:direct-derivation} are pushouts (cf. 
  Corollary~\ref{corollary:dd-po}). We state this corollary as follows:%
\end{isamarkuptext}\isamarkuptrue%
\isacommand{corollary}\isamarkupfalse%
\ \isanewline
\ \ \ \ \ \ {\isacartoucheopen}pushout{\isacharunderscore}{\kern0pt}diagram\ K\ L\ d{\isachardot}{\kern0pt}D\ G\ idM\ d{\isachardot}{\kern0pt}d\ g\ d{\isachardot}{\kern0pt}c{\isacharprime}{\kern0pt}{\isacartoucheclose}\ \isakeyword{and}\ {\isacartoucheopen}pushout{\isacharunderscore}{\kern0pt}diagram\ K\ R\ d{\isachardot}{\kern0pt}D\ g{\isachardot}{\kern0pt}H\ idM\ g\ g{\isachardot}{\kern0pt}h\ g{\isachardot}{\kern0pt}c{\isacartoucheclose}\ \isanewline
\isadelimproof
\ \ %
\endisadelimproof
\isatagproof
\isacommand{using}\isamarkupfalse%
\ \isanewline
\ \ \ \ d{\isachardot}{\kern0pt}po{\isachardot}{\kern0pt}pushout{\isacharunderscore}{\kern0pt}diagram{\isacharunderscore}{\kern0pt}axioms\isanewline
\ \ \ \ g{\isachardot}{\kern0pt}po{\isachardot}{\kern0pt}pushout{\isacharunderscore}{\kern0pt}diagram{\isacharunderscore}{\kern0pt}axioms\isanewline
\ \ \isacommand{by}\isamarkupfalse%
\ simp{\isacharunderscore}{\kern0pt}all%
\endisatagproof
{\isafoldproof}%
\isadelimproof
\endisadelimproof
\begin{isamarkuptext}%
Isabelle's simplifier is able to discharge the proof obligations after supplying the corresponding
  facts.
  In the upcoming section we state our conclusion and areas of future work.%
\end{isamarkuptext}\isamarkuptrue%
\isadelimdocument
\endisadelimdocument
\isatagdocument
\isamarkupsection{Conclusion and Future Work \label{sec:conclusion}%
}
\isamarkuptrue%
\endisatagdocument
{\isafolddocument}%
\isadelimdocument
\endisadelimdocument
\begin{isamarkuptext}%
Formal verification increases the trustworthiness and reliability of software. 
  In this paper, we present a revised version of \cite{Soeldner-Plump22a} on our formalisation
  of the double-pushout approach with injective matching over (node and edge) labelled directed 
  graphs, in the proof assistant Isabelle/HOL.
  The formalisation uses the extensible locale mechanism, which allows us to combine theories
  and to structure our work. Compared to earlier work in \cite{Soeldner-Plump22a}, we rely
  on total functions and follow Noschinski's approach of representing graphs and morphisms.
  With these changes, we reduced the required lines of text by 50 percent while enhancing overall
  readability.

  We first formalise graphs and morphisms and prove several properties, such as the 
  well-definedness of morphism composition. Direct derivations are introduced in terms of 
  deletion and gluing. We prove their correspondence to pushouts. In addition, we prove that 
  pushouts are unique up to isomorphism. Although, Isabelle is not able to discharge most of the 
  generated proof obligations automatically, the available proof methods support the discharging 
  process.

  Our formalisation of pushouts results in pushout objects of a to specific type, which prevents
  us from constructing an elegant proof within the deletion context (as we highlight at the end
  of Subsection~\ref{sec:deletion-pushouts}).
  In the meantime, we have overcome this limitation by imposing the constraint that nodes and edges
  must be natural numbers (only within the universal property). This is realised by a type synonym
  such as \isa{\isacommand{type{\isacharunderscore}{\kern0pt}synonym}\ {\isacharparenleft}{\kern0pt}{\isacharprime}{\kern0pt}l{\isacharcomma}{\kern0pt}{\isacharprime}{\kern0pt}m{\isacharparenright}{\kern0pt}\ ngraph\ {\isacharequal}{\kern0pt}\ {\isachardoublequoteopen}{\isacharparenleft}{\kern0pt}nat{\isacharcomma}{\kern0pt}nat{\isacharcomma}{\kern0pt}{\isacharprime}{\kern0pt}l{\isacharcomma}{\kern0pt}{\isacharprime}{\kern0pt}m{\isacharparenright}{\kern0pt}\ pre{\isacharunderscore}{\kern0pt}graph{\isachardoublequoteclose}}.
  By restricting our locale type variables to have an instance of the \isa{countable} class, we
  could define functions to convert between the natural and generic representation. 
  If \isa{{\isacharprime}{\kern0pt}a} is an instance of the \isa{countable} class (denoted by \isa{{\isacharprime}{\kern0pt}a\ {\isacharcolon}{\kern0pt}{\isacharcolon}{\kern0pt}\ countable}),
  there exists an injective function, mapping each element of type \isa{{\isacharprime}{\kern0pt}a} to an element of typ \isa{nat}.

  Future developments will also include the proof of classical DPO results such as the Church-Rosser
  theorem. We also plan the extension towards attributed DPO graph transformation.
  Our long-term goal is the development of a practical Isabelle-based proof assistant for the 
  verification of individual programs in the graph programming language 
  GP\,2 \cite{Campbell-Courtehoute-Plump21a}.%
\end{isamarkuptext}\isamarkuptrue%
\isadelimtheory
\endisadelimtheory
\isatagtheory
\endisatagtheory
{\isafoldtheory}%
\isadelimtheory
\endisadelimtheory
\end{isabellebody}%

\bibliographystyle{eptcs}
\bibliography{root}

\end{document}